\documentclass[final,12pt]{elsarticle}

\usepackage{amssymb}
\biboptions{comma,square}

\usepackage[utf8]{inputenc}
\usepackage{graphicx,url}

\journal{Physica A}

\begin{document}
\begin{frontmatter}



\title{Density outbursts in a food web model with closed nutrient cycle.}


\author{Janusz Szwabiński}
\ead{janusz.szwabinski@ift.uni.wroc.pl}
\address{Institute of Theoretical Physics, University of Wrocław, pl. M. Borna 9, 50-204 Wrocław, Poland}

\begin{abstract}
A spatial three level food web model with a closed nutrient cycle is presented and analyzed via Monte Carlo simulations. The time evolution of the model reveals two asymptotic states: an absorbing one with all species being extinct, and a coexisting one, in which concentrations of all species are non-zero. There are two possible ways for the system to reach the absorbing state. In some cases the densities increase very quickly at the beginning of a simulation and then decline slowly and almost monotonically. In others, well pronounced peaks in the $R$, $C$ and $D$ densities appear regularly before the extinction. Those peaks correspond to density outbursts (waves) traveling through the system. We investigate the mechanisms leading to the waves. In particular, we show that the percolation of the detritus (i. e. the accumulation of nutrients) is necessary  for the emergence of the waves. Moreover, our results corroborate the hypothesis that top-level predators play an essential role in maintaining the stability of a food web (top-down control).  

\end{abstract}

\begin{keyword}

Monte Carlo simulations \sep Food webs \sep Food web viability \sep Predator-prey systems \sep Nutrient cycle \sep Density outbursts \sep Traveling waves \sep Percolation
\end{keyword}

\end{frontmatter}


\section{Introduction}
\label{intro}

Every year significant efforts are directed to conservation and restoration of natural ecosystems. In order to develop efficient protection strategies of those systems identification and understanding of mechanisms that lend them their stability are required.

Starting with the pioneering work of Elton~\cite{elton27}, food webs are a central organizing theme in studying ecosystems. They depict feeding relationships among species in ecological ecosystems. Since their introduction by Elton, they have been subject of intensive studies from both theoretical~\cite{morin,wil2000,drossel,amaral,PSBD} and experimental~\cite{morin,deangelis} perspective.

Although recent decades have seen a significant progress in understanding of food web properties, our knowledge remains sketchy. Until 1970s the predominant belief of ecologists was that large and highly complex ecosystems were more stable that the simpler ones~\cite{elton27,arthur55}. It was then questioned by May~\cite{may72}, who showed mathematically that large and complex random communities are inherently unstable. May's have initiated a debate on stability of ecosystems which lasts to this day and is still far from being over.

Several mechanisms have been identified as factors leading to stability of food webs. One of them is compartmentalization, i.e. existence of subsets of species in the food web that interact more frequently among themselves than with other species in the system. It was shown~\cite{stouffer2011} that compartments are benefitial to an ecosystem, because they act to buffer the propagation of extinctions. A special case of compartments are groups of consumers associated with each particular plant species, called component communities~\cite{thebault2010}. A food web consisting of such communities is stable, because a disturbance associated with fluctuactions in species' density is confined largely to that species' component community. 

Generalist consumers constitute another stabilizing factor~\cite{thompson2007}. A generalist is able to switch from one food source to another, which is more abundant. This switching tends to keep a food web stable, since it allows to control the abundant species and lets the less common one recover~\cite{neutel2007}.  

A process that often occurs in food webs is a top-down control of lower trophic levels by apex predators~\cite{estes2011}. The top-level predators play a key role in controlling the population of prey and, as a consequence, they limit the degree to which the prey endanger primary producers. Many population collapses may be traced down to altered top-down forcing regimes associated with the loss of native apex predators or the introduction of exotic species.

Detritus consisting mostly of bodies or fragments of dead organisms has been long recognized as one of the important factors in ecology. However, the theories of food webs and trophic dynamics have largely neglected detritus-based chains and have focused almost exclusively on
grazing food chains~\cite{moore}. The ‘‘green-world’’ view of these works has been severely criticized~\cite{polis} and it is expected that
only merging this approach with the one considering also the  detritus could yiels a satisfactory ecological theory~\cite{moore}.
The existing theoretical studies~\cite{polis97} indicate that the flow of energy from detritus to living chain may increase the extinction cascades. There are also arguments~\cite{mccann} that consumption of prey from detritus chain may weaken top-down regulations of stability. Thus, many fundamental questions asked in ecology, concerning the structure of food webs, the length of food chains, the size and direction of extinction cascades, may have a different form and interpretation when detritus is taken into account.

Food webs consist of species using dispersal as a mean of  both procuring key resources and avoiding natural enemies. Therefore they are inherently spatial entities, a fact which is also neglected in the vast majority of studies leading to a rather poor understanding of spatial food web dynamics~\cite{ama2008}. An ubiquitous feature of different spatial systems in nature are waves traveling through them. Those waves are of big importance, because they  control the speed of many dynamical processes including chemical reactions~\cite{kur03}, epidemic outbreaks~\cite{gre01} and biological evolution~\cite{sny03}. Although great effort has been expended into the description of the waves~\cite{sny03,hal11}, their rigorous description is still lacking mainly due to their sensitivity to different kinds of fluctuations. Thus, any contribution that provides insight into the mechanisms governing the traveling waves is valuable to our understanding of complex systems with spatial structures. 

In our recent study~\cite{jsw10,szw12a} a three species food web model with a detritus path was analysed by means of Monte Carlo simulations. Our findings indicate that under certain conditions complex spatio-temporal patterns in form of density waves appear in the food web. Analysis of those waves is the main goal of this work.

The paper is organized as follows. In Section~\ref{model} the model is introduced. Simulation results are discussed in Section~\ref{results}. And finally in Section~\ref{conclusions} conclusions are drawn.

\section{Model}
\label{model}

\begin{figure}
\centering
\includegraphics[scale=0.5]{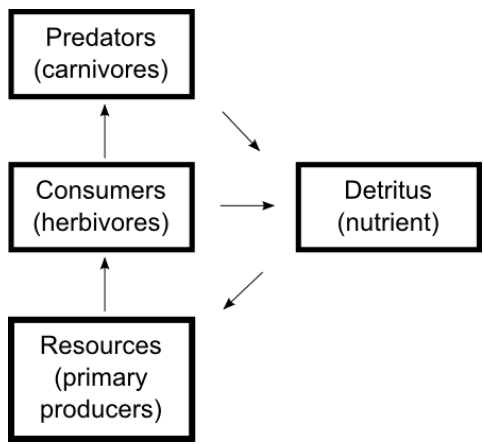}
\caption{Model food web investigated in this paper. The arrows indicate the flow of nutrient in the trophic loop. Consumers $C$ feed on the resources $R$ and are themselves food for predators $P$. Detritus $D$ consisting of dead consumers and/or predators provides nutrient for resources. 
\label{fig: food web 2}}
\end{figure}

Fig.~\ref{fig: food web 2} illustrates a hypothetical system we are going to investigate. The food web consists of three trophic levels. The basal level species will be called \textit{resources} ($R$) within our model. It corresponds to primary producers or autotrophs in real ecosystems, i.e. to organisms able to take energy from the environment (sunlight or inorganic chemicals) and convert it into energy-rich molecules such as carbohydrates. 

The species at intermediate level in Fig.~\ref{fig: food web 2} will be called \textit{consumers} ($C$). It relates to herbivores in real systems, which principally feed on primary producers.

The consumers themselves constitute food for the top level species called \textit{predators} ($P$). In general, the predators correspond to carnivores in real systems.

The remains of the consumers and predators form detritus ($D$), which provides nutrient for the resources.

The model food web was built up under the tacit assumption that the conversion of dead fragments into nutrient occurs immediately and without any external help. In reality, dead organisms are broke down and converted into useful chemical products by decomposers. For the sake of simplicity we will neglect them in the present work and address their role for the viability of the whole food web in a forthcoming paper.

It should be noted that the model shown in Fig.~\ref{fig: food web 2} resembles some real marine food webs with phytoplankton being the primary producer, zoo plankton and fish occupying the higher trophic levels and bacteria playing the role of decomposers~\cite{fishegg}.

We are going to investigate the model by making use of an agent-based Monte Carlo simulation. To this end we put individuals of each species on a square lattice. Each individual is characterized by two parameters: a death rate $d_i$ and a birth rate $b_i$. The death rate determines the probability that an organisms dies in a given time step if it is not able to feed. The birth rate stands for the ability of an individual to convert food reserves into breeding success.
Theoretically, at each trophic level there may live different species. We will assume however that each level is occupied by only one species. Moreover, to keep the model as simple as possible, we do not differentiate between organisms of a given species. In other words, each individual of the species $i$ ($i=R,C$ or $P$) will have exactly the same pair of the parameters $(b_i,d_i)$.

Once put randomly on the lattice, the system will evolve due to the following rules~\cite{jsw10,szw12a}: 
\begin{enumerate}
\item While populating the nodes we are applying an exclusion principle stating that each lattice node may be occupied by at most one agent of a given type. Multiple occupancy of nodes by individuals of different species is allowed.
\item If a $C$ appears at a node populated by an $R$, then $R$ is eaten and removed from the lattice. This may correspond to a herbivore eating a plant. Next, a progeny of $C$ is created with a probability $b_C$ at a randomly chosen node in its Moore neighborhood, provided there is no other $C$ at the new node yet. The rule reflects the fact that in the real life there is usually a close correspondence between food reserves of an individual and its breeding success.
\item If a $P$ appears at a node occupied by a $C$, an analogous situation takes place. $P$ feeds on $C$ and then produces an offspring with a probability $b_P$. 
\item If there is no $R$ at a node occupied by a $C$, then the consumer dies with a probability $d_C$. Its body $D$ stays at the node until used by an $R$.
\item Similarly, if there is no $C$ at a node populated by a $P$, the predator dies with a probability $d_P$ and turns into a detritus portion $D$.
\item If an $R$ meets a $D$ at a node, it feeds  and then produces offspring at empty nodes in its Moore neighborhood. The eaten detritus is removed from the lattice. While the upper level species $C$ and $P$ are allowed to produce at most one progeny after each breeding success, for the basal species $R$ we assume that it may populate all empty sites (i.e. 8 nodes at most) in the neighborhood after breading, each of which with the probability $b_R$.  This assumption complies with the fact that the lower the trophic level the higher its productivity~\cite{odum}.
\item If there is no $D$ at a node occupied by an $R$, it dies with a probability $d_R$ and is removed from the lattice. Note that there is no conversion into a $D$ in this case.
\end{enumerate}

\begin{table}
\centering
\begin{tabular}{lll}
\hline
State in time $t$ &  & State in time $t+1$ \\ 
\hline $E$ & $\longrightarrow $ & $E$ \\ 
$R$ & $\longrightarrow $ & $R$ or $E$ \\ 
$C$ & $\longrightarrow $ & $C$ or $D$ \\ 
$RC$ & $\longrightarrow $ & $C$ ($+C$) \\ 
$P$ & $\longrightarrow $ & $P$ or $D$ \\ 
$D$ & $\longrightarrow $ & $D$ \\ 
$RP$ & $\longrightarrow $  & $RP$ or $RD$ or $P$ \\ 
$CP$ & $\longrightarrow $ & $P$ ($+P$) \\ 
$RCP$ & $\longrightarrow $ & $P$ ($+P$) \\ 
$RD$ & $\longrightarrow $ & $R$ ($+$ max $8R$) \\ 
$CD$ & $\longrightarrow $ & $CD$ or $D$ \\ 
$RCD$ & $\longrightarrow $ & $CD$ \\ 
$PD$ & $\longrightarrow $ & $PD$ or $D$\\ 
$RPD$ & $\longrightarrow $ & $RP$ ($+$ max $8R$) \\ 
$CPD$ & $\longrightarrow $ & $PD$ ($+P$)\\ 
$RCPD$ & $\longrightarrow $ & $RP$ ($+P$,$+$ max $8R$) \\
 \hline 
\end{tabular} 
\caption{Look up table in our model. $E$, $R$, $C$, $P$ and $D$ stand for empty node, resource, consumer, predator and dead individual (consumer or predator), respectively. The values in parentheses represent a possible outcome of an action in the neighborhood of a given site.\label{look-up}}
\end{table}

As it follows from the above rules, a single node of the lattice may be in one of 16  states. Depending of its actual state different actions may be performed according to the rules. Possible states and outcomes of those actions are summarized in Table~\ref{look-up}. Looking at this table one can immediately see that the system reveals a very rich dynamics. That is why we made some further assumptions. First of all, we take only occurrences of $D$ into account, not its actual quantity. As a consequence, if a consumer or a predator dies at a node already containing a $D$, then its content will remain one $D$. In other words, we simply assume that resources are able to absorb all detritus they find. Thus, its quantity does not really matter. A further presumption is that the dead resources do not contribute to the nutrient pool.

A living species in our model does not move on the lattice, it remains localized. The only way of invading new lattice sites is via proliferation. In the simplest version of the model the directions of proliferation are chosen at random. However, it is possible to put some intelligence into the behavior of the species.

The impact of different proliferation strategies was already discussed in Ref.~\cite{szw12a}. In that work the following strategies has been taken into account:
\begin{itemize}
\item $Random$ - a mixture of the randomness and exclusion, i.e. the aforementioned basic rules.
\item $CtR$ - consumers $C$ put their progeny only on a site occupied by $R$, i.e. the direction of their invasion is driven by food availability. If there is no $R$ in the neighborhood of $C$, no offspring is produced. Other species follow the $Random$ strategy.
\item $CaP$ - consumers $C$ put their progeny on a site not occupied by $P$. Other species follow the $Random$ strategy.
\item $CtRaP$ - this is a mixture of the above strategies, i.e. consumers $C$ put their progeny on a site occupied by $R$, if its not occupied by $P$ at the same time. Other species follow the $Random$ strategy.
\item $PtC$ - predators $P$ put their progeny only on a site occupied by $C$. Other species follow the $Random$ strategy.
\item $CaPtC$ - predators $P$ put their progeny only on a site occupied by $C$, whereas $C$ avoid sites occupied by $P$. Resources $R$ follow the $Random$ strategy.
\end{itemize}

If not stated otherwise, the results presented in the next sections of this paper have been obtained with the $Random$ strategy. However we will refer to other strategies at many points for the sake of comparison.

\section{Results}
\label{results}

With the assumptions mentioned in the previous section, the model has seven important parameters: birth and death rates of each species $(b_i,d_i)$ ($i=R,C,P$) and the linear size of the lattice $L$. To reduce the number of parameters we arbitrarily fix all death rates at $0.01$.

We set the linear size $L$ of the lattice equal to 200 throughout the simulations. It was already shown in Ref.~\cite{jsw10} that for system sizes larger than 200 the viability of the food web depends only weakly on $L$. Thus, $L=200$ constitutes a reasonable compromise between the computational efforts and quality of the results.

All results presented in this section were obtained in Monte Carlo simulations with the Moore neighborhood (i.e. 8 neighboring sites) on the square lattice, because it turned out to be a good compromise between the richness of the resulting dynamics of the system and the computational requirements. Among other choices we have tried the von Neumann neighborhood with only 4 neighboring sites and extended Moore-like ones with 12 and 20 sites. While the von Neumann neighborhood led to an absorbing state (see Section~\ref{asymptotics} for more details) independently of the actual model parameters, the extended neighborhoods yielded results which were qualitatively the same compared to the less computational demanding Moore case.

Most simulations were performed up to $10^5$ Monte Carlo steps (MCS). This particular value turned out to be a reasonable choice as well, because changing it to higher values did not significantly modified the results. In other words, it is highly unlikely in our model that a system, which is alive at $T=10^5$, will die out afterwards.

If required (e.g. for phase diagrams in Sec.~\ref{phase diagrams}), we averaged the results over 100 independent runs in order to get a good statistics.      

\subsection{Asymptotic states}
\label{asymptotics}

Since the asymptotic states of the system have been already discussed in Refs.~\cite{jsw10} and~\cite{szw12a}, we give here only a brief summary of the most important findings.

In case of proliferation strategies $Random$, $CtR$ and $PtC$ (see Sec.~\ref{model} for definitions) the system ends up in one of two distinct asymptotic states depending on the particular values of the model parameters: in an absorbing one with all species being extinct or in a coexisting one, in which all species survive till the end of a simulation.

Theoretically, one could expect that a state with species $R$ and $C$ having non-zero concentrations and $P$ being extinct is possible as well. Such a state was not observed in the simulations. Thus the predators seem to be indispensable for the survival of the system.

Three other strategies considered here, i.e. $CtRaP$, $CaP$ and $CaPtC$, end up in an absorbing state independently of the actual model parameters. The reason is as follows. The presence of the predators in the system is required to control the number of consumers. Otherwise, consumer population will grow rapidly and kill all available resources driving the entire system to extinction. According to Table~\ref{look-up}, only nodes in the states $CP$, $RCP$, $CPD$ and $RCPD$ can lead to the migration of the predators, which is needed to maintain their population. In general, there are three sources of those four states: (1) states resulting from initial conditions, (2) predators which put their progeny on nodes occupied by C's and (3) consumers which put their offspring on nodes occupied by P's. In the strategies that end up in a coexisting state, all those sources are present. However,  in the $CtRaP$, $CaP$ and $CaPtC$ strategies consumers avoid nodes occupied by predators, i.e. the third source is missing~\cite{szw12a}. Although predators from initial $CP$ and $RCP$ states (source 1) are able to proliferate or even migrate farther if there are some consumers in their neighborhoods (source 2), there is no way for them to proceed with proliferation after all consumers in their vicinity will be eaten up because no $P$ node will be populated by a new $C$. As a consequence, they will die out after a while. 

Closer look at the absorbing states reveals that there are two different ways for the system to reach them. Depending on the parameters of the model, in some cases the concentrations of species decrease almost monotonically towards zero. Other possibility is that well pronounced peaks in the $R$, $C$ and $D$ densities appear before extinction.

\begin{figure}
\centering
\includegraphics[scale=0.4]{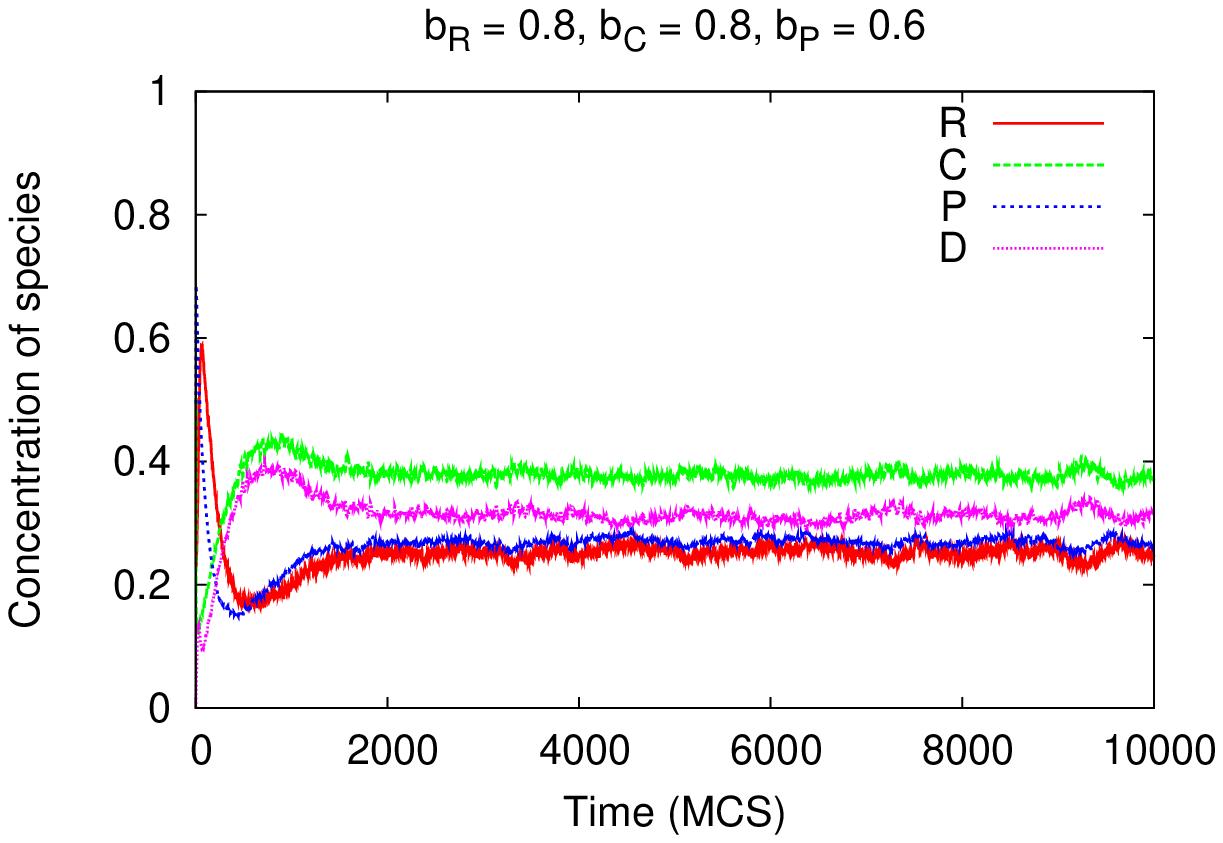}\\
\includegraphics[scale=0.4]{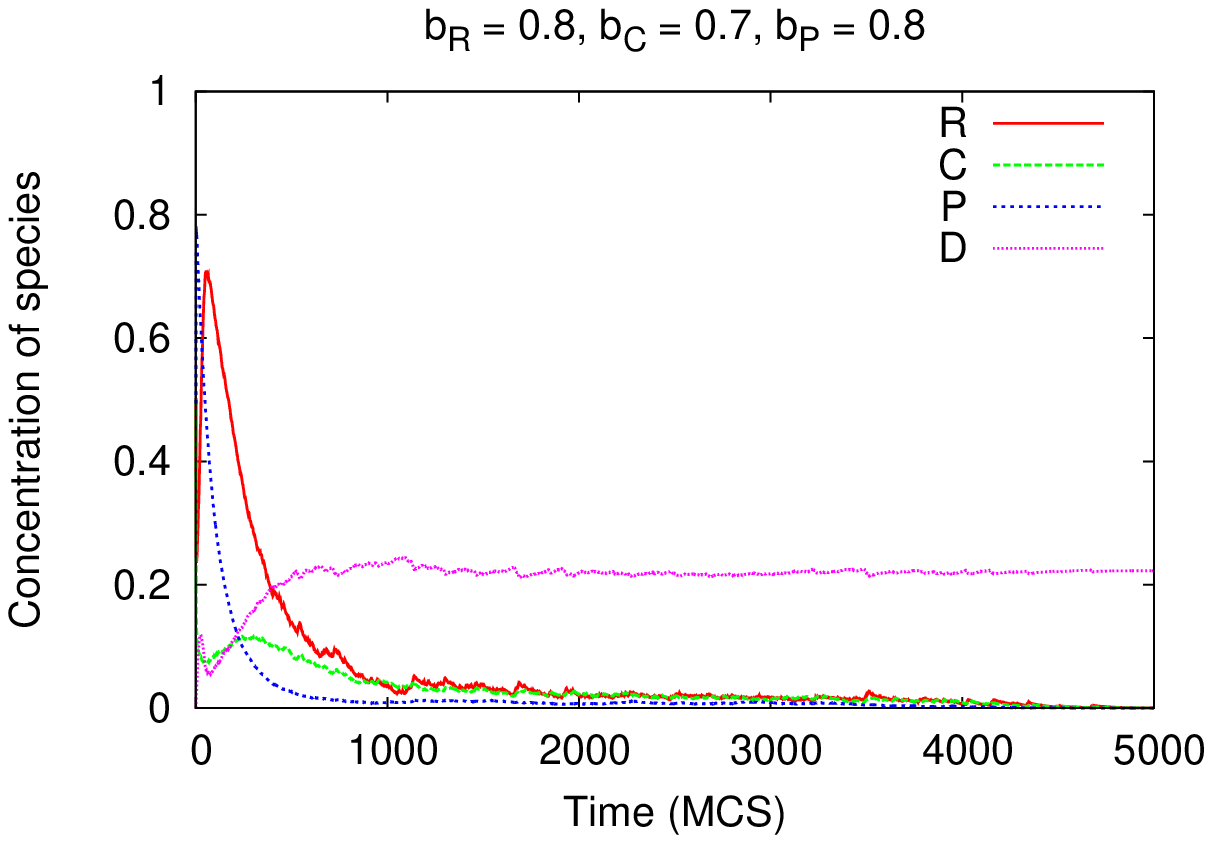}\
\includegraphics[scale=0.4]{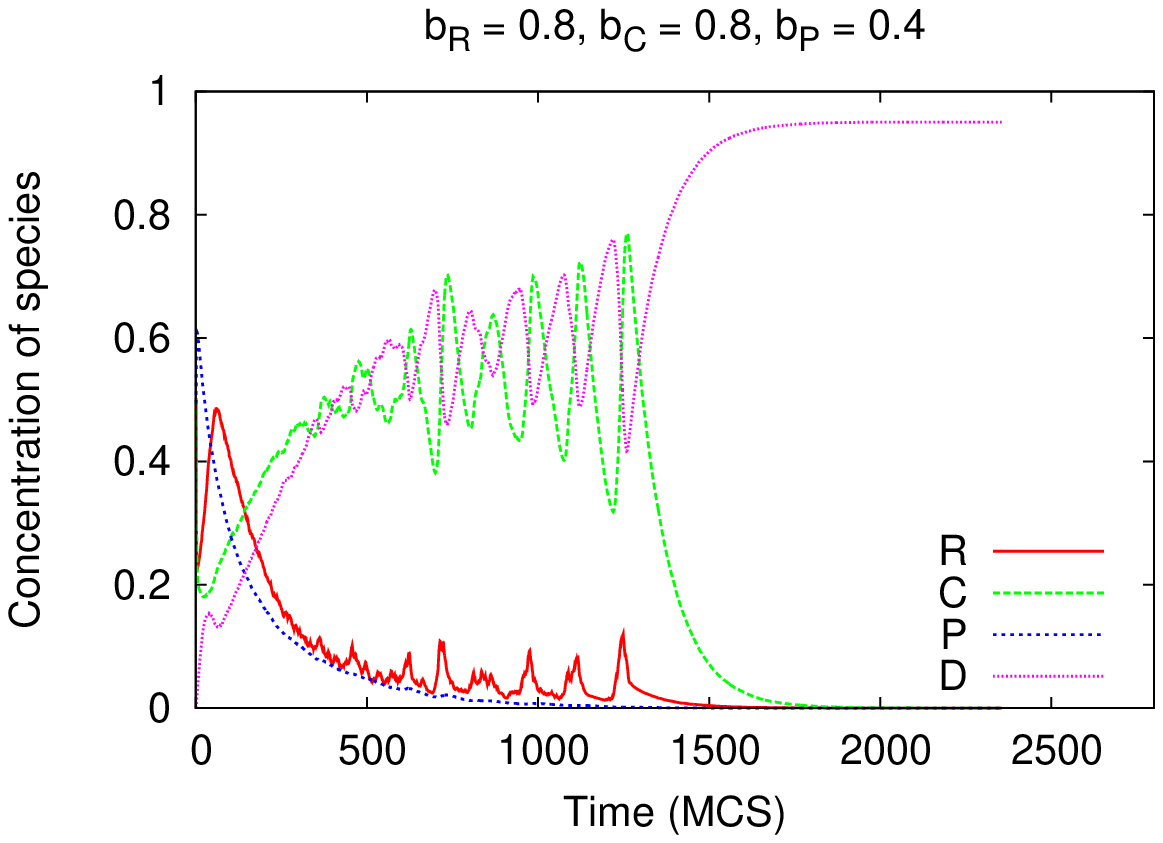}
\caption{(Color online) Example of asymptotic states in case of the $Random$ strategy. In the alive state (top plot) all species survive till the end of the simulation. In the absorbing state (bottom plots) the entire system dies out. The are two distinct ways for the system to reach the absorbing state: almost monotonic decrease of the species densities (bottom left) and well pronounced peaks of the densities before the extinction (bottom right). Birth rates used in the simulations are indicated in the plots. Death rates were all set to 0.01 and the linear size of the lattice was equal to 200. \label{fig: asymptotics}}
\end{figure}

Examples of each asymptotic state in case of $Random$ strategy are shown in Fig.~\ref{fig: asymptotics}. The curves were obtained from single runs. The parameters of the simulations may be found in the plots and their captions.

We will show that the peaks seen in the bottom right plot in Fig.~\ref{fig: asymptotics} correspond to density waves traveling through the system. Analysis of those waves is the main goal of this paper.

\subsection{Origin of density outbursts}

\begin{figure}
\centering
\includegraphics[scale=0.4]{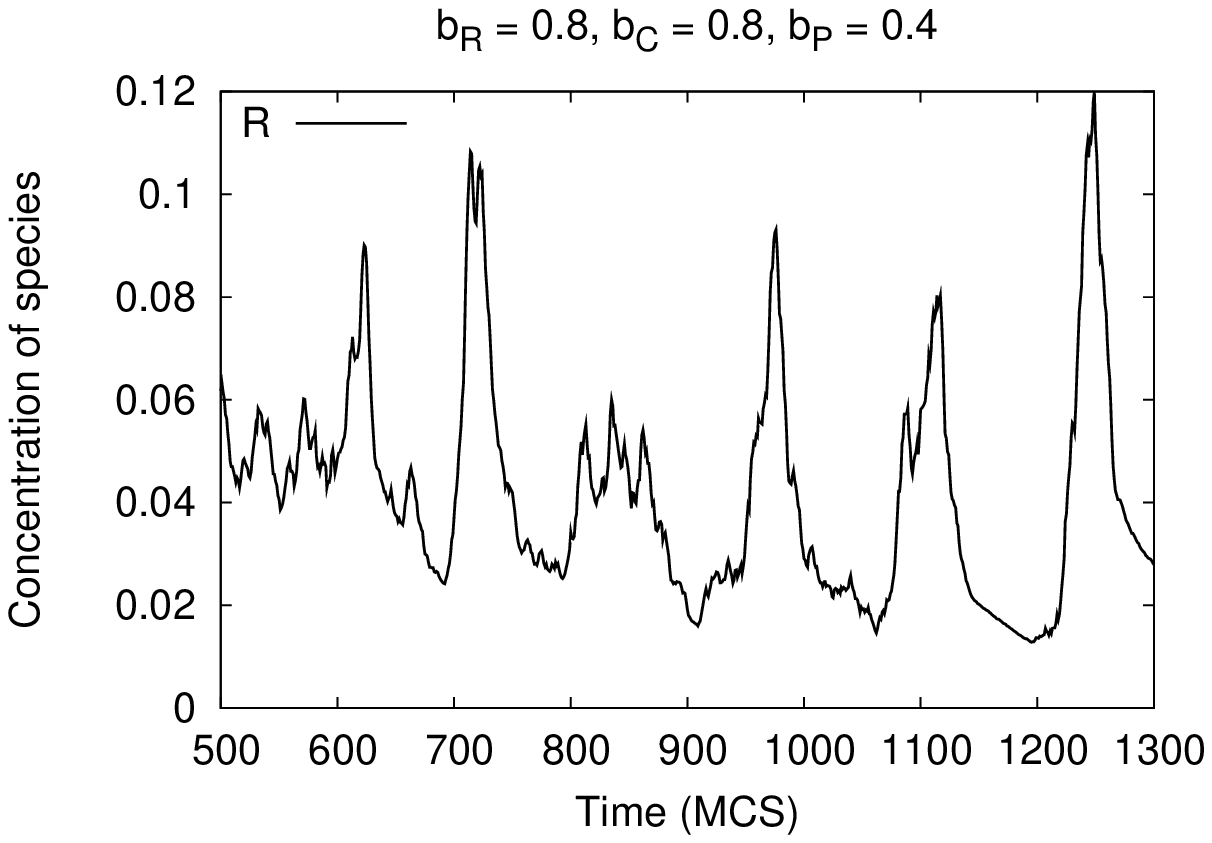}\
\includegraphics[scale=0.4]{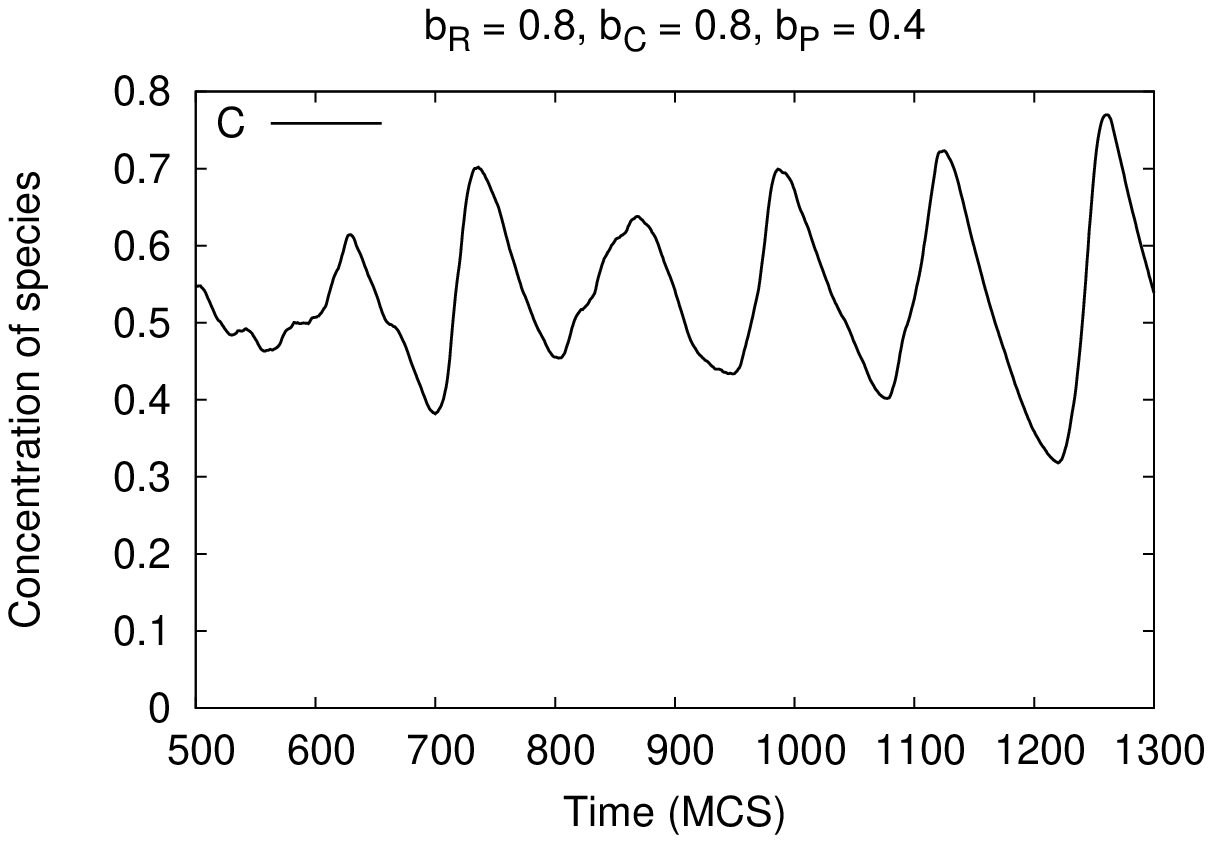}
\caption{Well pronounced peaks in the densities of $R$ (left plot) and $C$ (right plot). See bottom right plot in Fig.~\ref{fig: asymptotics} for more details. \label{fig: asymptotics 2}}
\end{figure}

In Fig.~\ref{fig: asymptotics 2} the time evolution of $R$ and $C$ in the system from the bottom right plot of Fig~\ref{fig: asymptotics} is presented again in a smaller time window. As far as the resources $R$ are concerned, we observe rapid explosions of their population followed by almost equally rapid declines. This behavior bears a resemblance to characteristics of an excitable media~\cite{mur90}. Moreover, a similar time evolution may be found in processes that take place in plankton populations and are known as "spring blooms" and "red tide" phenomena~\cite{tru94}.

\begin{figure}
\centering
\includegraphics[scale=0.5]{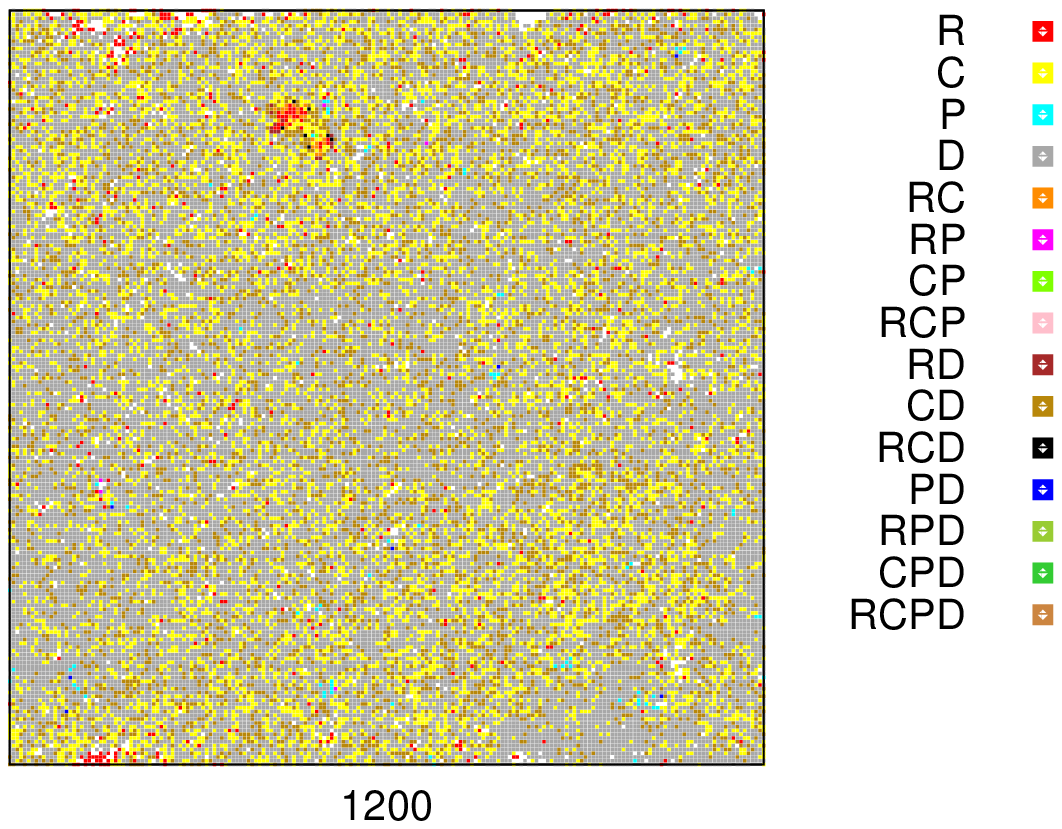}
\includegraphics[scale=0.5]{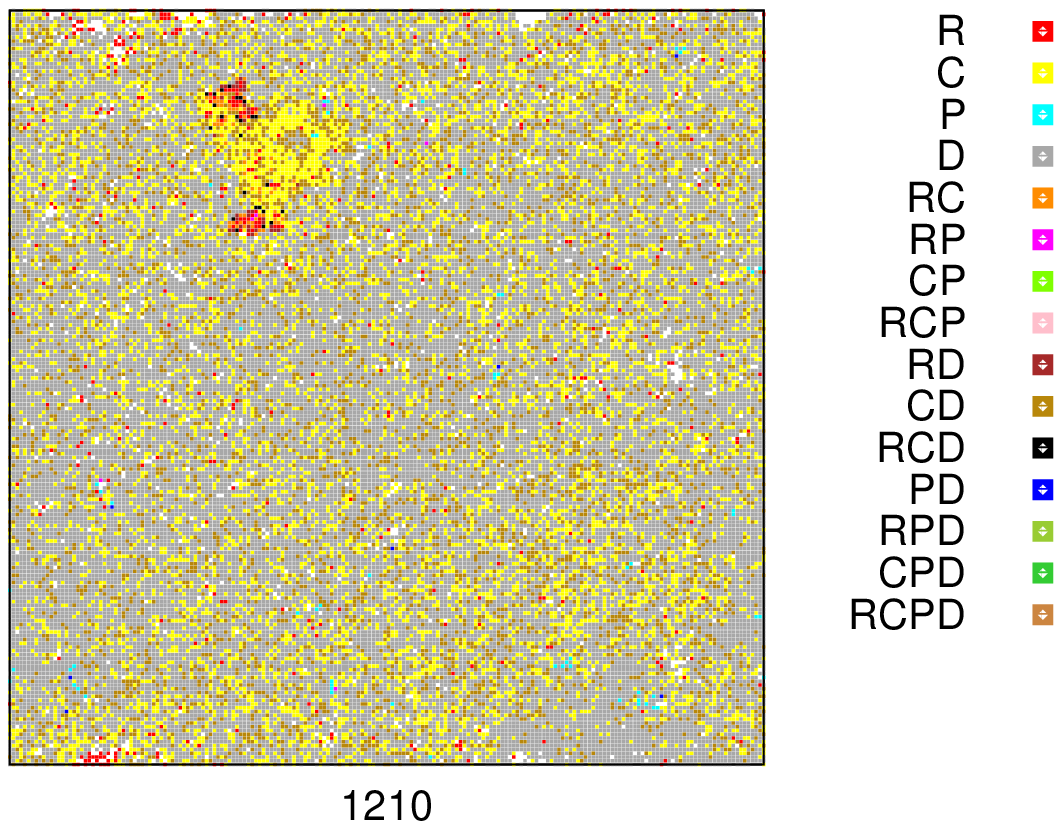}\\
\includegraphics[scale=0.5]{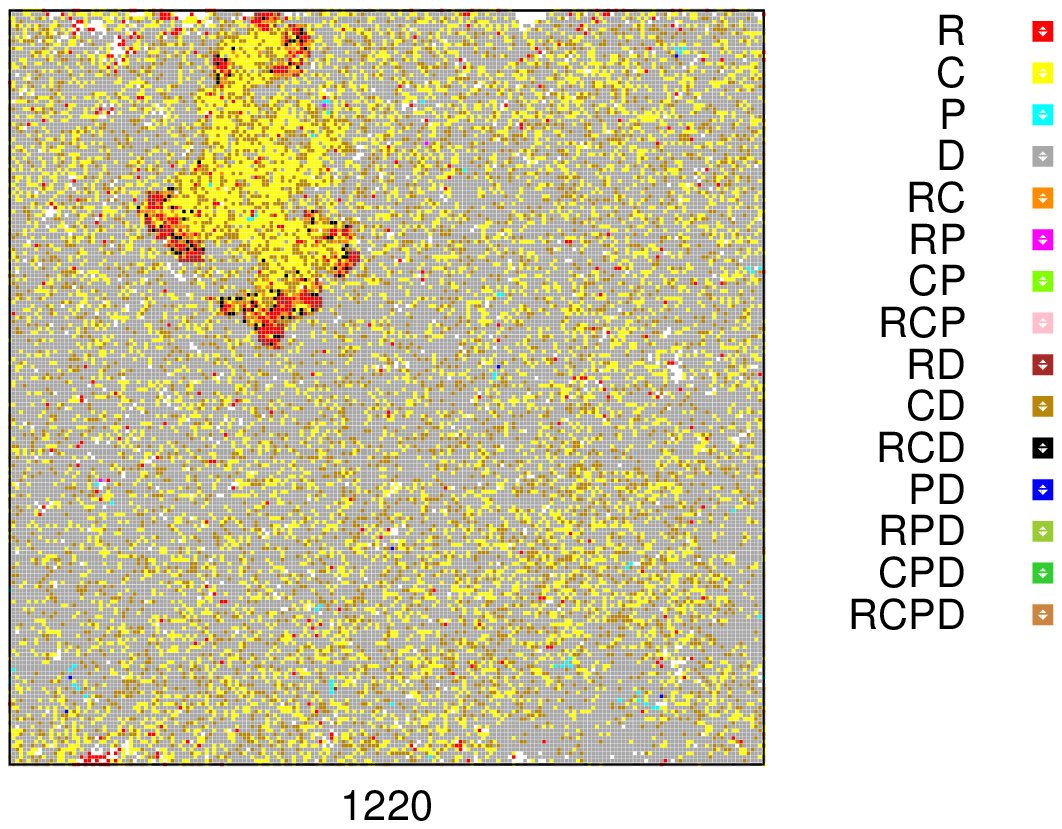}
\includegraphics[scale=0.5]{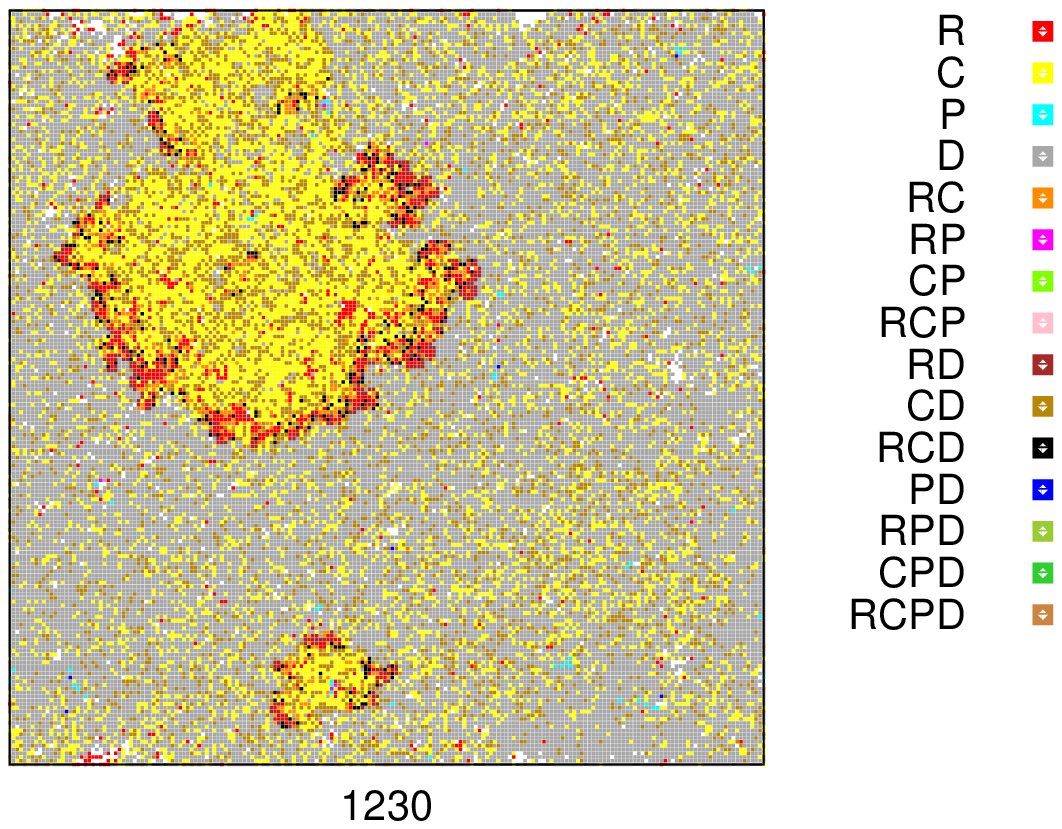}\\
\includegraphics[scale=0.5]{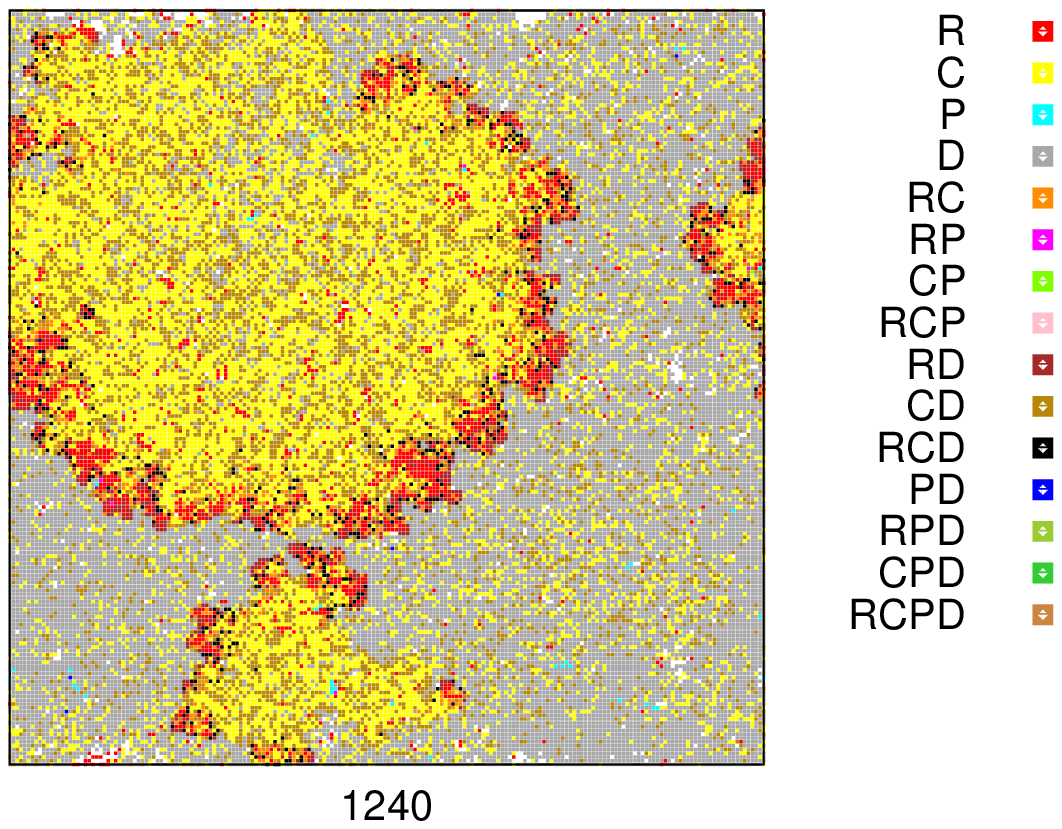}
\includegraphics[scale=0.5]{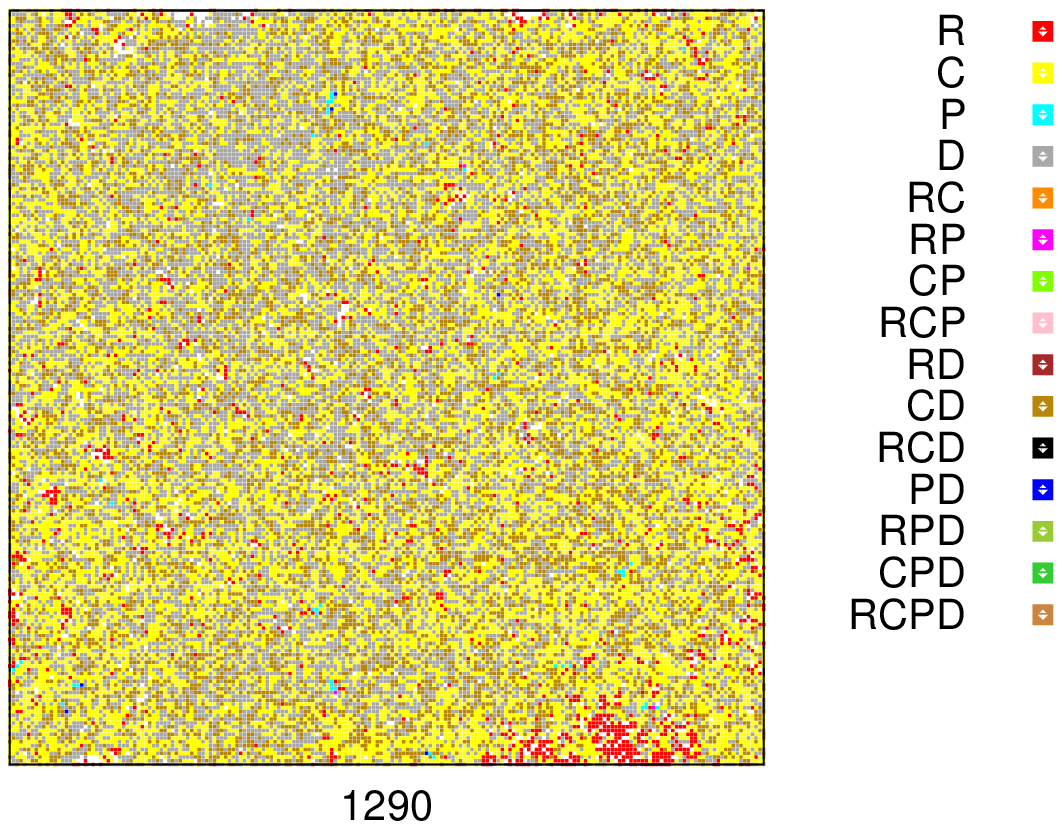}
\caption{(Color online) Snapshots of the system at different MC steps. At $t=1200$ MCS, most of the nodes are in a $C$ (yellow) or $D$ (grey) state. There is only a small cluster of $R$ (red) in the top center part of the lattice. This cluster grows, splits into two parts and moves away from its origin, as it may be seen in the snapshot at $t=1210$ MCS. It is followed by even bigger cluster of $C$'s. Both clusters continue to grow and spread over the entire system. Finally the resources hit the boundaries of the system (or the front of another wave) and have no place to further escape from $C$'s. They will be diminished very quickly by the consumer's wave. Then the consumers will partially die out due to the lack of food and again, most of the nodes will be in a $C$ or $D$ state. Parameters of the simulation: $L=200$,$b_R=0.8$,$b_C=0.8$,$b_P=0.4$.\label{fig: patterns}}
\end{figure}

Let us now have a closer look at the system at a microscopic level. Snapshots of the system at different MC steps are shown in Fig.~\ref{fig: patterns}. At $t=1200$ MCS, i.e. just before the last peak in the density of $R$ shown in Fig.~\ref{fig: asymptotics 2} occurs, there is almost no activity in the system. Most of the nodes are occupied by a $C$ or a $D$ and there are only a few resources and predators. However, there is a small cluster of $R$ in the top center part of the lattice. This cluster grows, splits into two parts and moves away from its origin, as it may be seen in the snapshot at $t=1210$ MCS. Note, that it is followed by an even bigger cluster of $C$'s. Both clusters continue to grow and spread over the entire system. This fact is reflected by the peaks in the densities as may be seen in Fig.~\ref{fig: asymptotics 2}. Finally the resources hit the boundaries of the system (or the front of another wave) and have no place to further escape from $C$'s. They will be diminished very quickly by the consumer's wave. Then the consumers will partially die out due to the lack of food - the densities of both species decline rapidly.  

\begin{figure}
\centering
\includegraphics[width=0.28\textwidth]{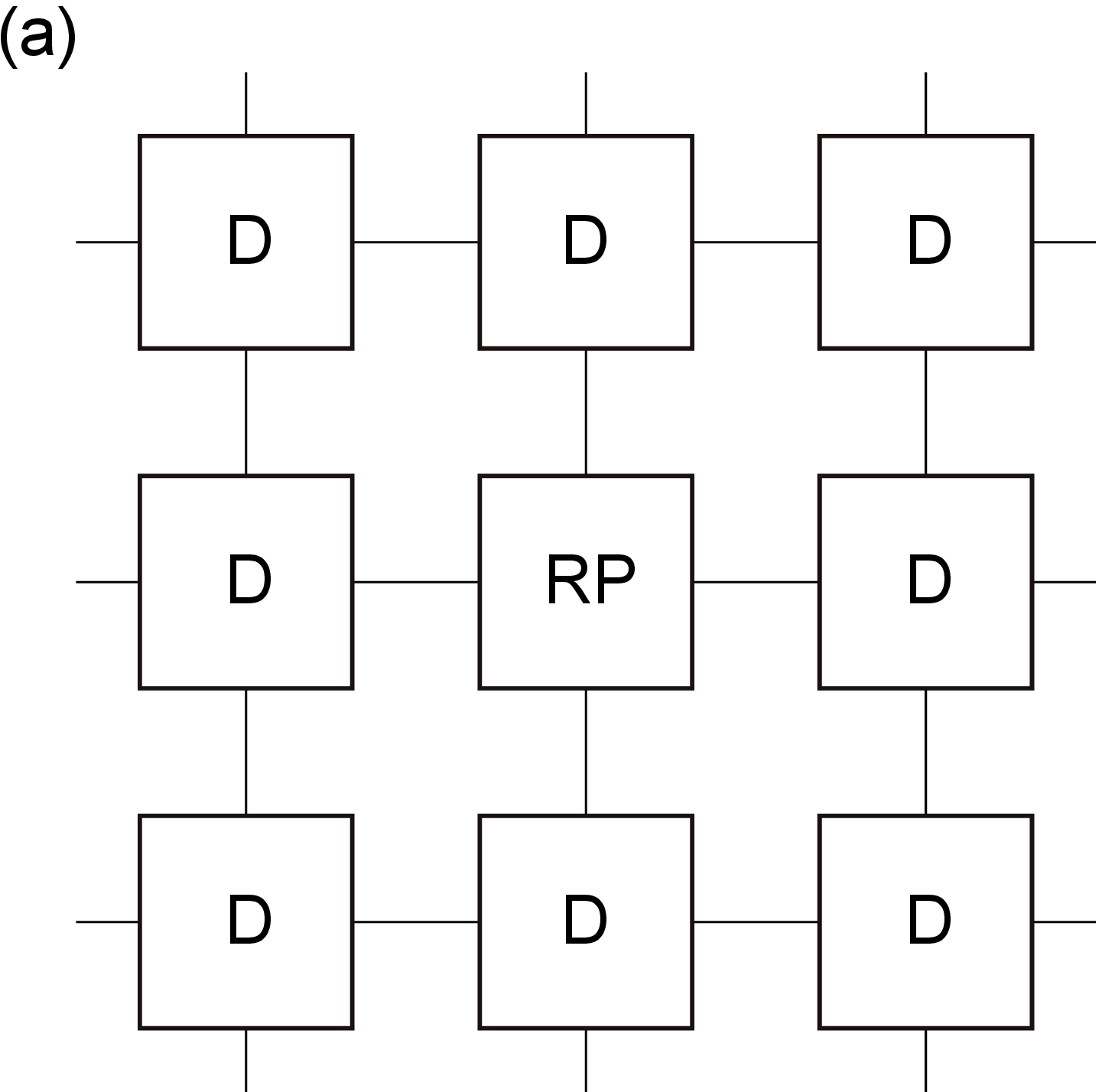}
\includegraphics[width=0.28\textwidth]{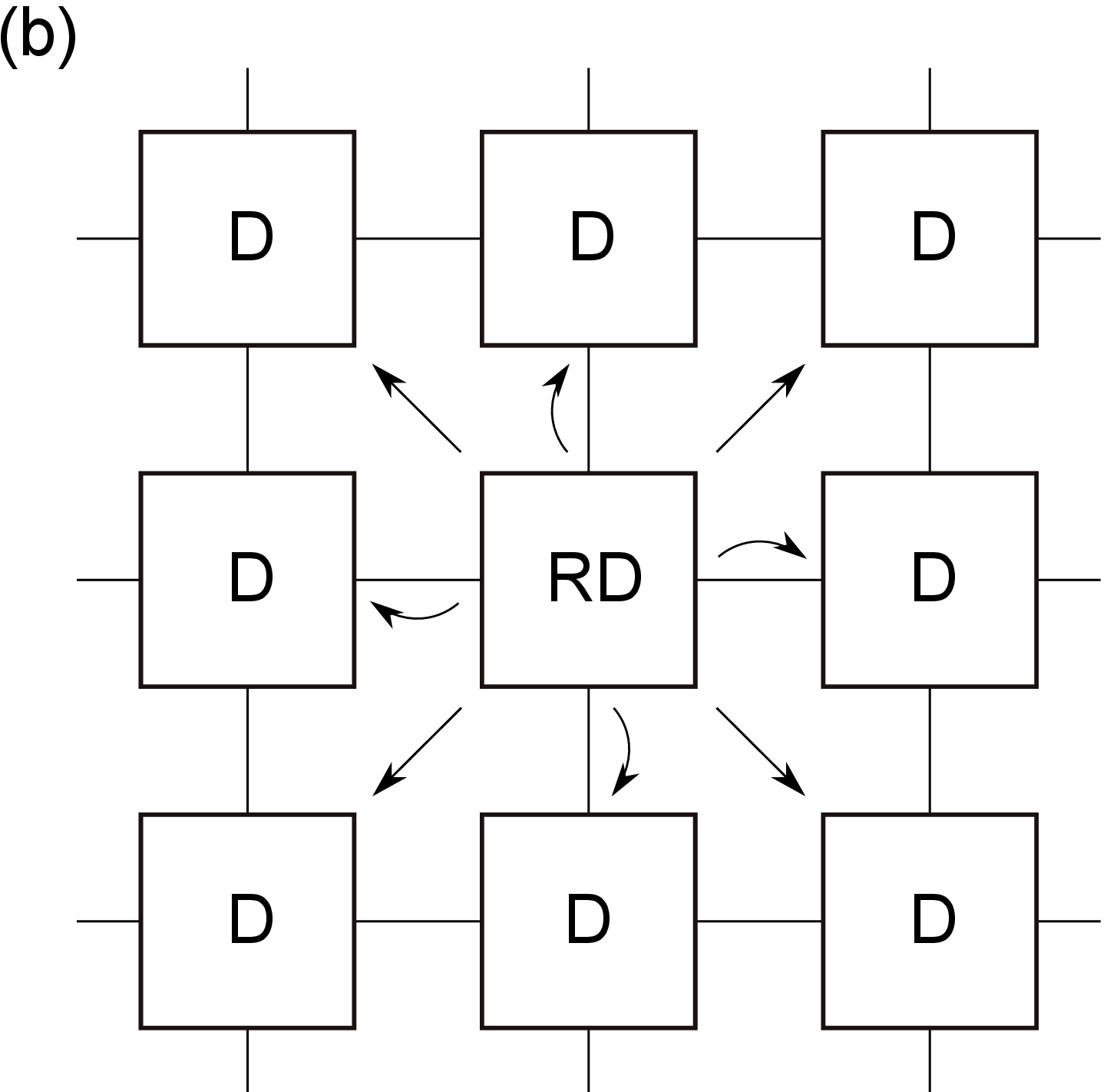}
\includegraphics[width=0.28\textwidth]{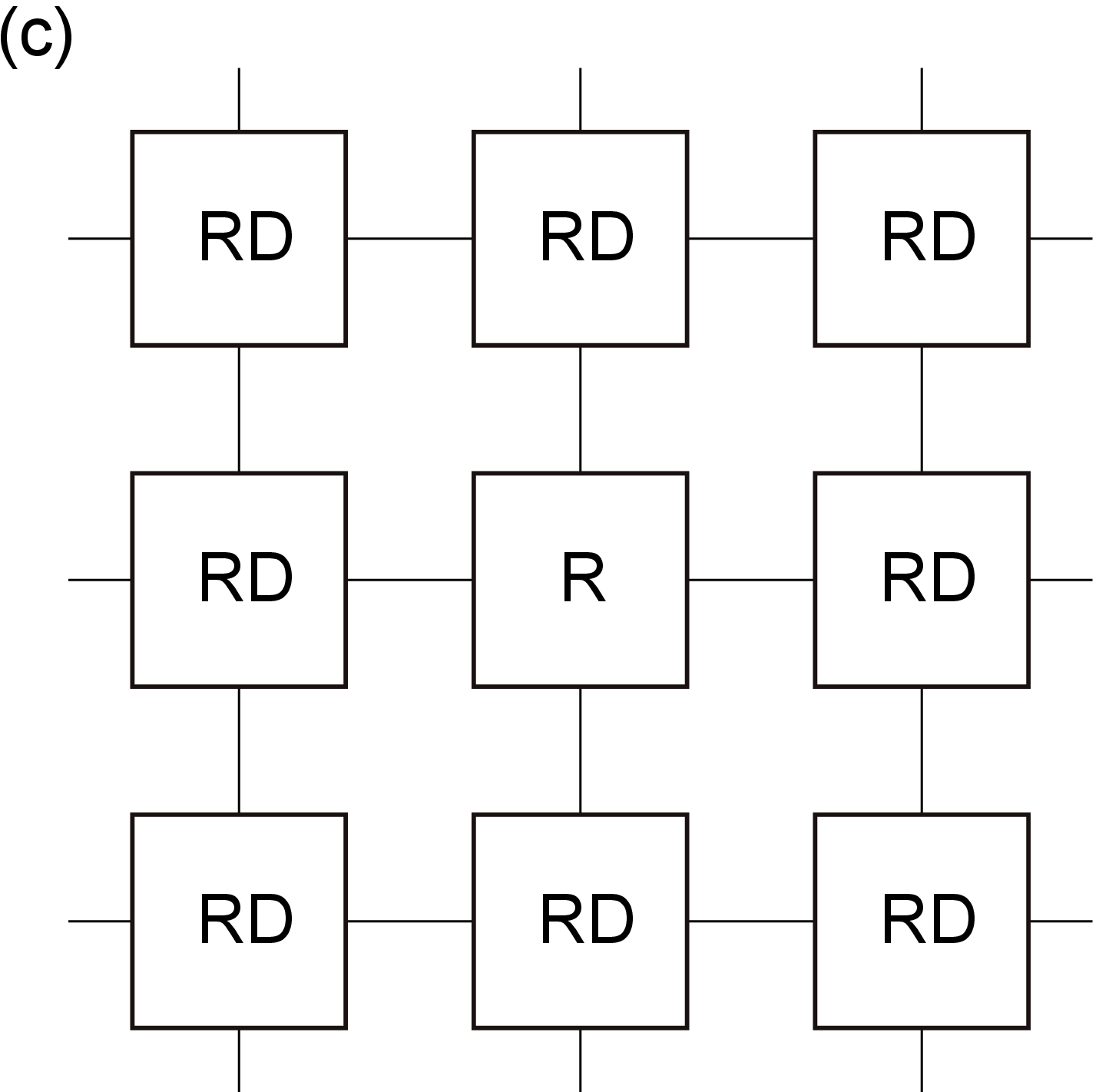}
\caption{Origin of $R$ blooms. A resource $R$ on an $RP$ node is protected by a predator $P$ from being eaten by a consumer. If $P$ dies, it provides food for $R$ and lets it proliferate. If the origin node is surrounded by the detritus, newborns get enough food to reproduce and an avalanche starts.\label{fig: rwaves}}
\end{figure}

Since the traveling waves presented above were observed only for some particular values of the simulation parameters, i.e. the birth rates of the species at different trophic levels, it is to expect that special conditions are necessary for an $R$ wave to emerge (see  Fig.~\ref{fig: rwaves}). First, there must be an $RP$ node, so that $R$ is protected from being eaten by a $C$ for some time interval. If $P$ dies due to the lack of food, it provides nutrient for $R$, which then proliferates to the neighboring sites. Second, if the origin $RP$ node was surrounded by $D$'s, the progenies are able to feed immediately and produce offspring by themselves. In this way the initial cluster seen in the top left snapshot in Fig.~\ref{fig: patterns} is created. Third, the overall concentration of $D$ must be high in order to feed the wave through the entire process.

\begin{figure}
\centering
\includegraphics[width=0.28\textwidth]{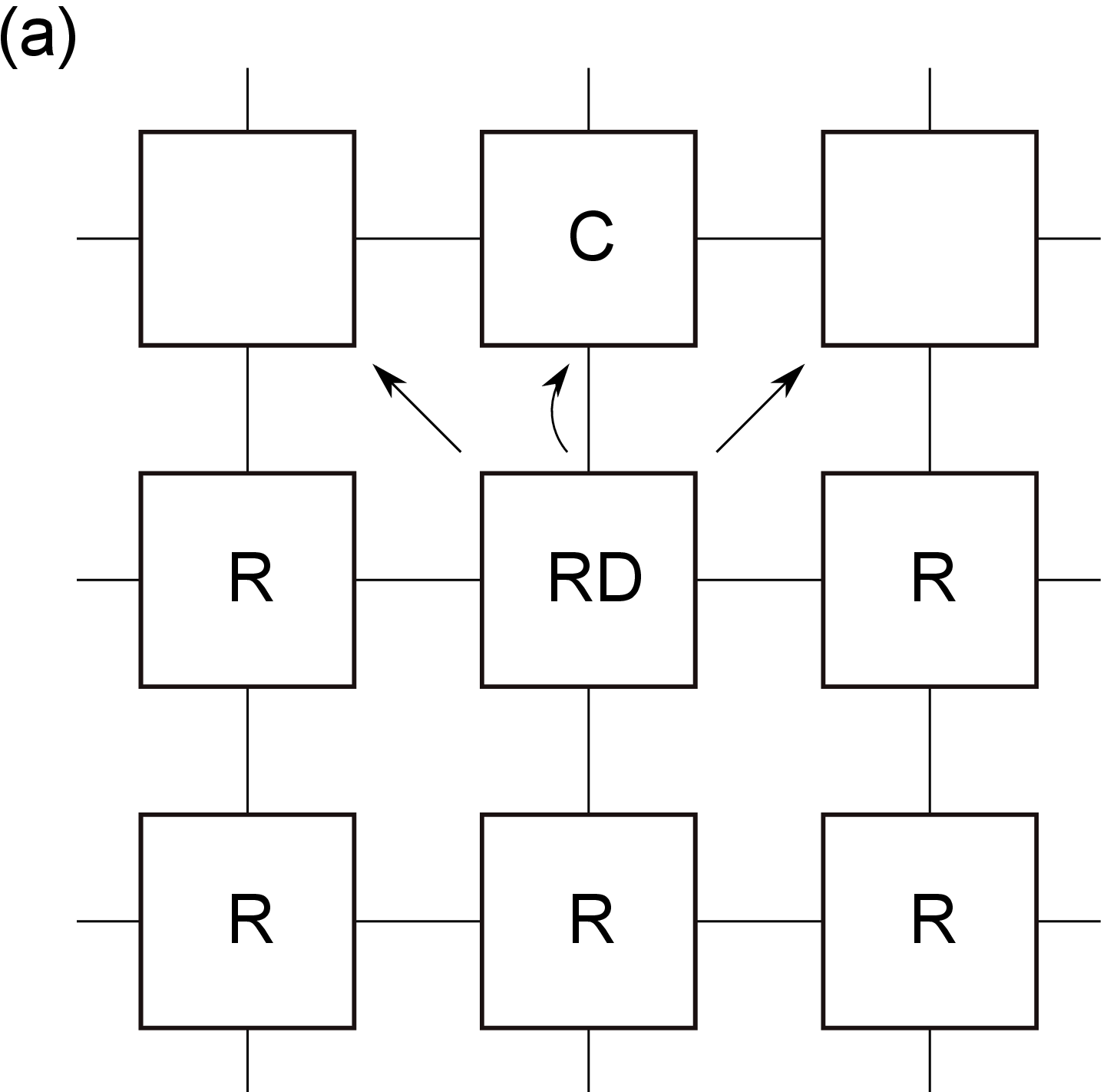}
\includegraphics[width=0.28\textwidth]{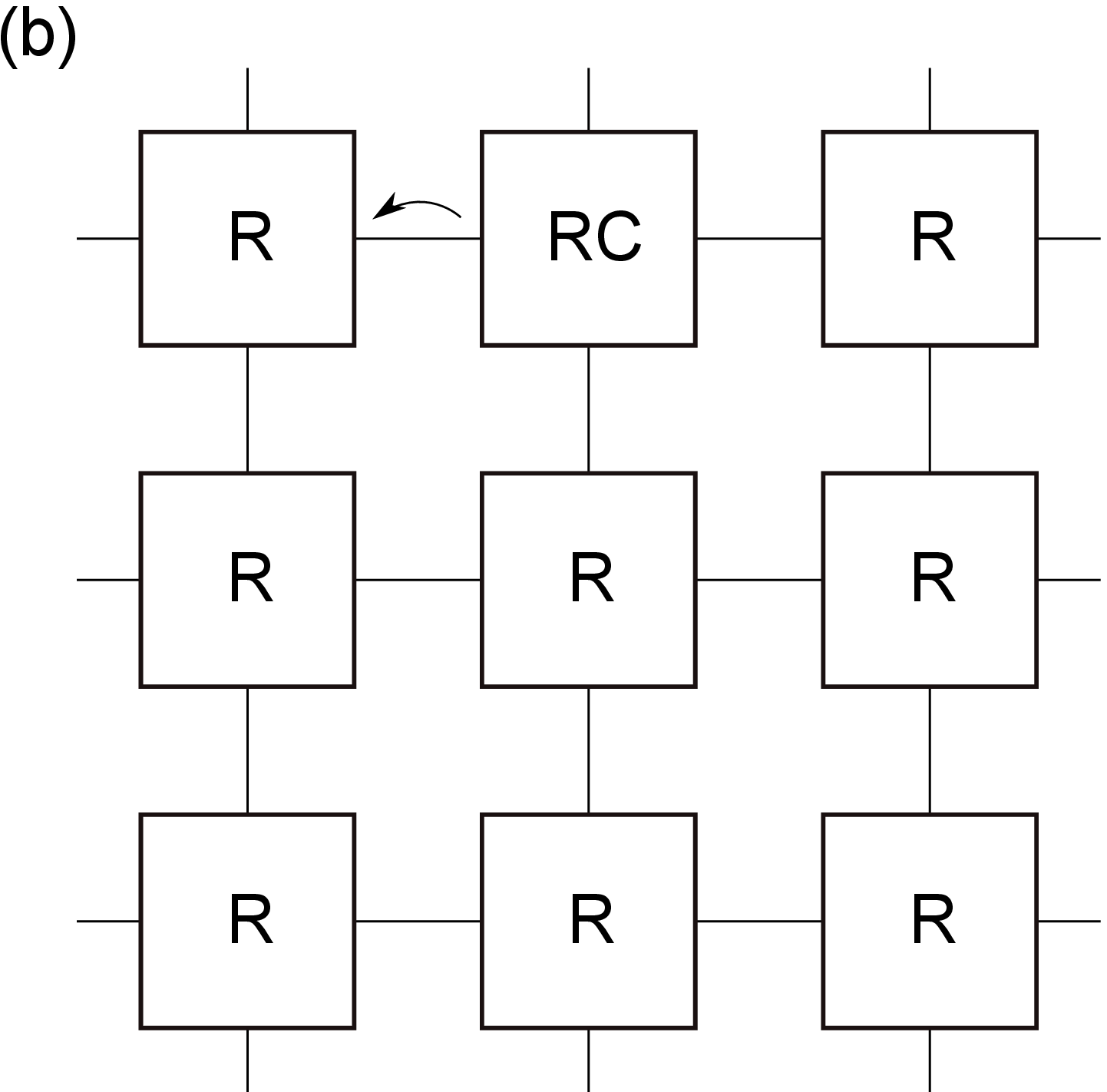}
\includegraphics[width=0.28\textwidth]{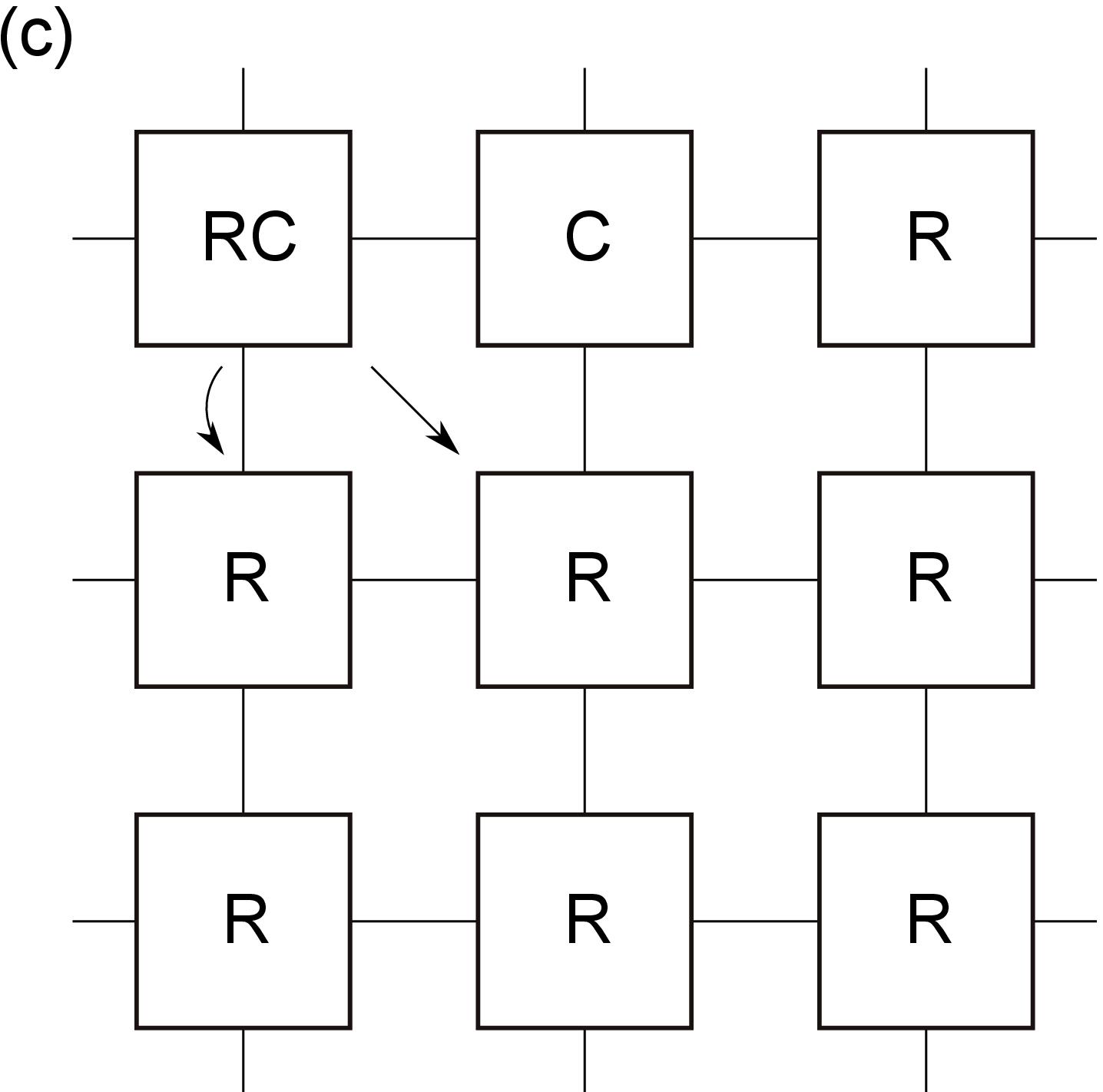}
\caption{Origin of $C$ blooms. If an $R$ wave hits a node populated by a $C$, it gets food and starts to produce offspring. Since there is plenty of food for $C$ newborns brought by the wave, they start to invade the lattice and a $C$ wave emerges. \label{fig: cwaves}}
\end{figure}

As may be seen from Fig.~\ref{fig: patterns}, the $R$ wave is followed by a $C$ bloom. An explanation for this phenomenon may be found in Fig.~\ref{fig: cwaves}. The $C$ avalanche is triggered by the front of an $R$ wave hitting a node occupied by a consumer. Since in all proliferation strategies under consideration the species $R$ follows a random behavior pattern, in the situation presented in Fig.~\ref{fig: cwaves} it will put an offspring on the $C$ node as well. As a result, the consumer gets food and starts to proliferate. Since there is plenty of food brought by the $R$ wave, the new consumers start to invade the lattice - a $C$ wave emerges. 

\subsection{Spanning clusters}

In the previous section it has been already mentioned that an $RP$ node and detritus states in its vicinity are necessary to feed the wave front (see Fig.~\ref{fig: rwaves}). Let us now investigate these conditions in more detail.

\begin{figure}
\centering
\includegraphics[scale=0.5]{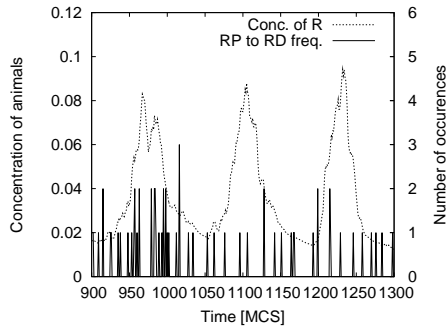} \
\caption{Frequency of $RP\rightarrow RD$ events overimposed on the resource density. Proliferation strategy: $Random$. Parameters of the simulation: $b_R=0.8$, $b_C=0.8$, $b_P=0.4$.\label{fig: rp freq}}
\end{figure}

We begin with checking the frequency of $RP\rightarrow RD$ events. As it follows from Fig.~\ref{fig: rp freq}, although their frequency decreases slightly in the course of time, such events occur in the low density phase not only just before the emergence of a new wave. Thus, an RP node surrounded by the detritus is not a sufficient condition for the wave.

\begin{figure}
\centering
\includegraphics[scale=0.45]{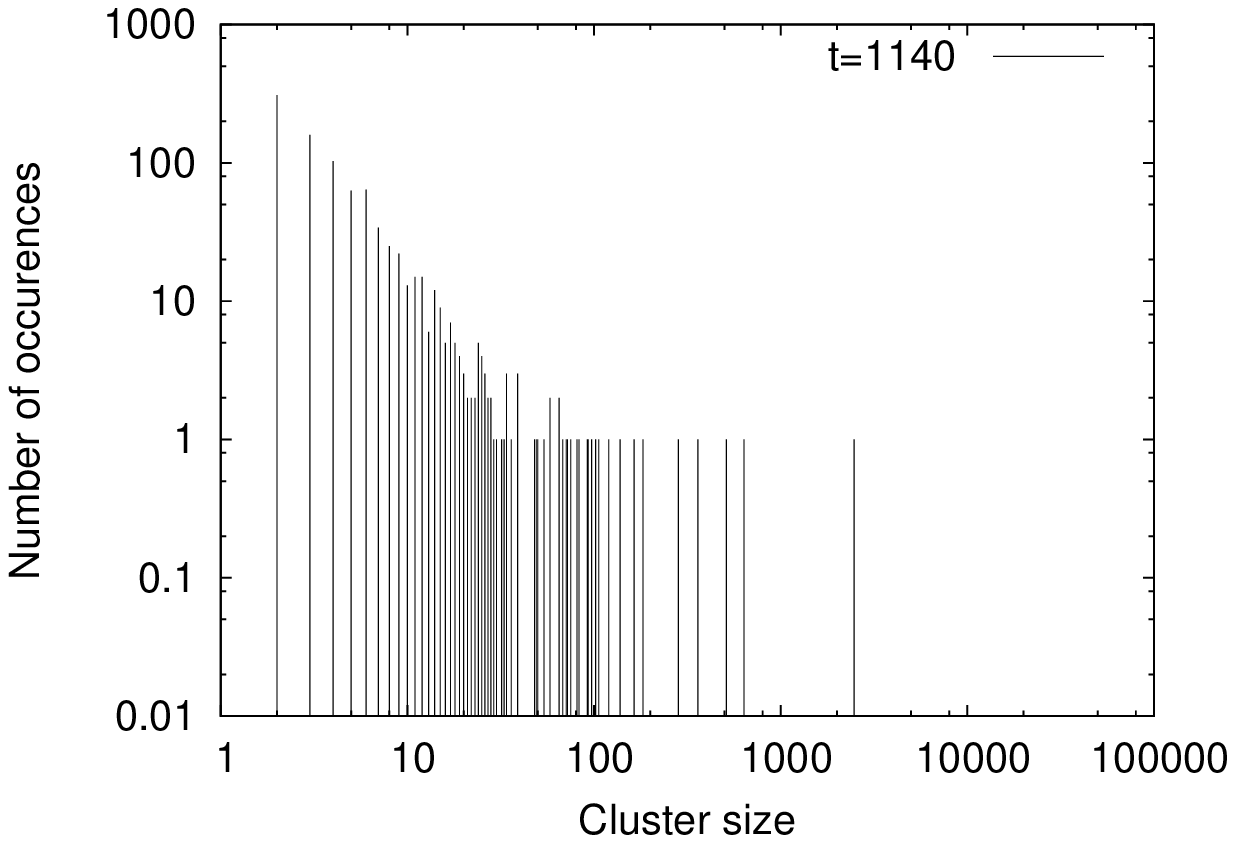} \
\includegraphics[scale=0.45]{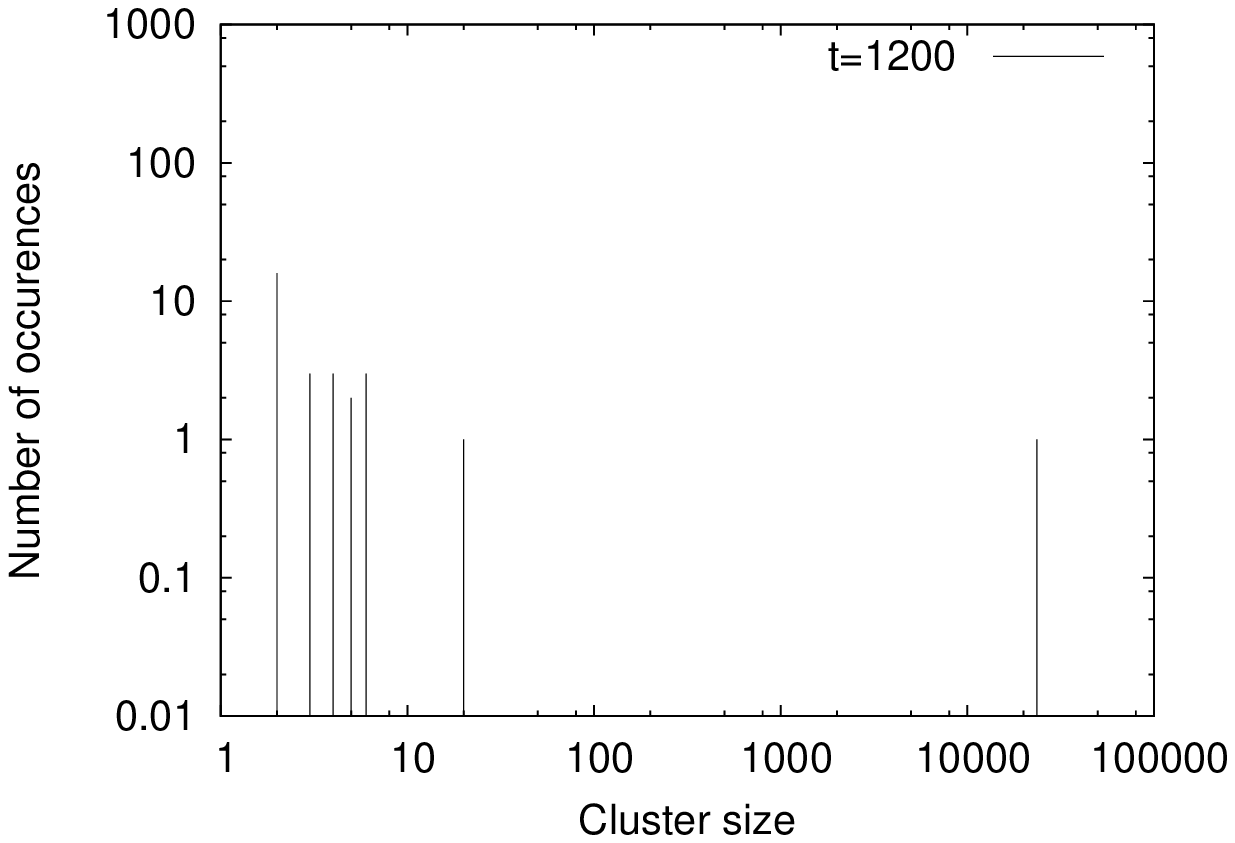} \
\includegraphics[scale=0.45]{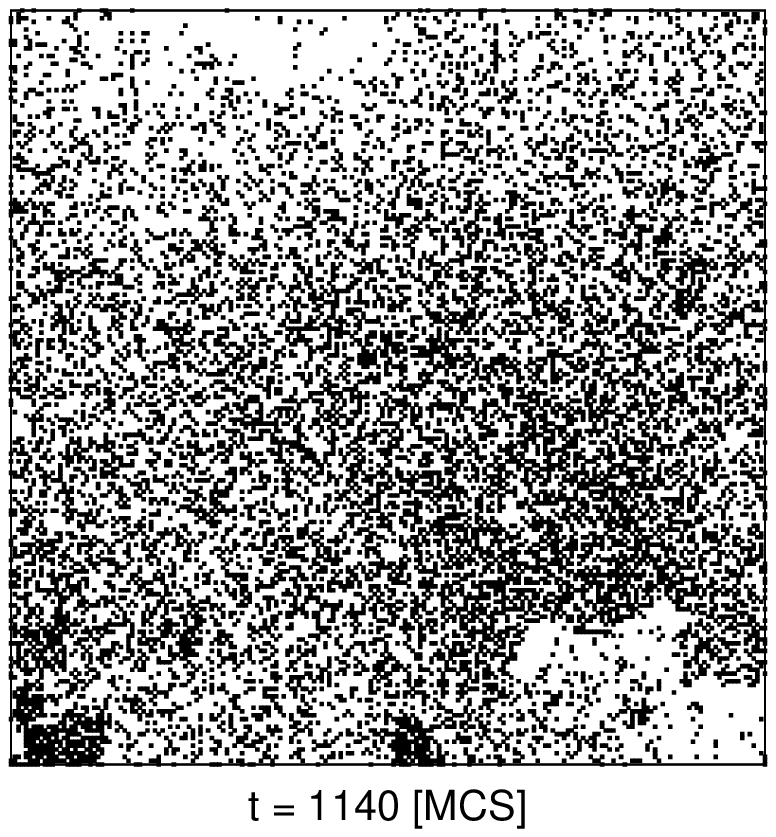} \
\includegraphics[scale=0.45]{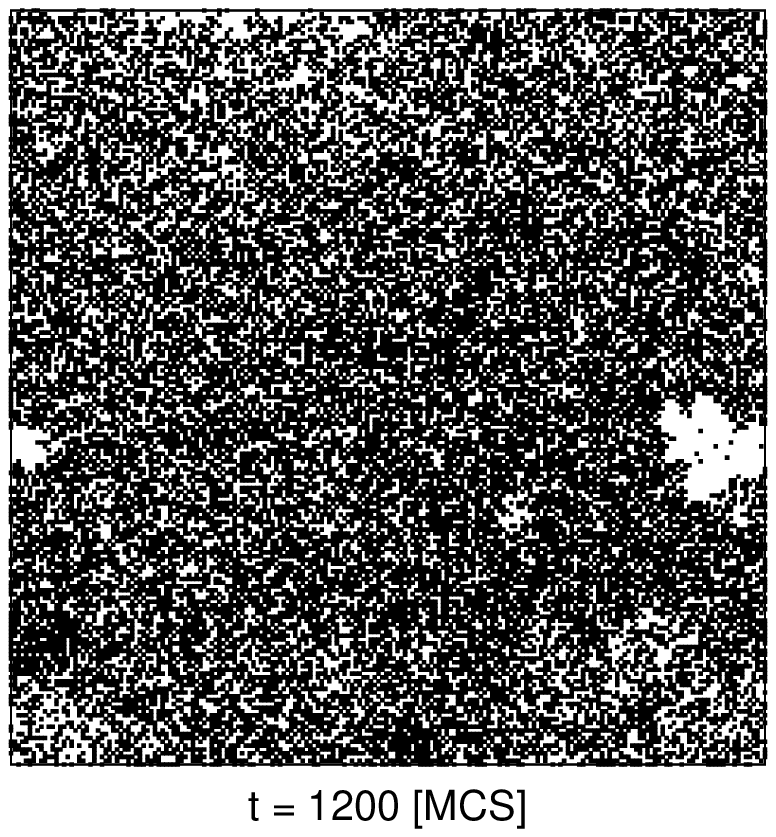} \
\caption{Top: Detritus cluster distributions at two different MC steps. Parameters of the simulation: $b_R=0.8$, $b_C=0.8$, $b_P=0.4$. At $t=1140$ MCS (left plot), there are many small clusters on the lattice. The cluster distribution at $t=1200$ MCS is different - there is one big cluster comparable with the size of the lattice and only few small ones. Bottom: Snapshots of $D$ clusters corresponding to the above distributions. \label{fig: d cluster dist}}
\end{figure}

In Fig.~\ref{fig: d cluster dist}, distributions of detritus clusters at two different MC steps are presented. Both distribution were measured in the low density state of the resources. However, while the left plot falls into the time interval just after the density decline, the right plot describes a phase before the formation of a new wave. We see that the two distributions differ significantly. At the beginning of the low density phase, there are plenty of small detritus clusters. If an $RP$ node is surrounded by such a small cluster, the emerging wave is suppressed sooner or later due to the lack of food for the $R$ newborns.
The situation changes upon entering the state corresponding to the right plot in  Fig.~\ref{fig: d cluster dist}. There are just a few small clusters and a big one with the size comparable with the size of the lattice. If an $R$ wave starts to form at the boundary or in the bulk of the big cluster, it can then travel through the entire system. Of course, the cluster itself will be destroyed by the wave and it will take a while to rebuild it. That is why not all $RP\rightarrow RD$ events lead to a density bloom and why we observe low density states in the system.

\begin{figure}
\centering
\includegraphics[scale=0.55]{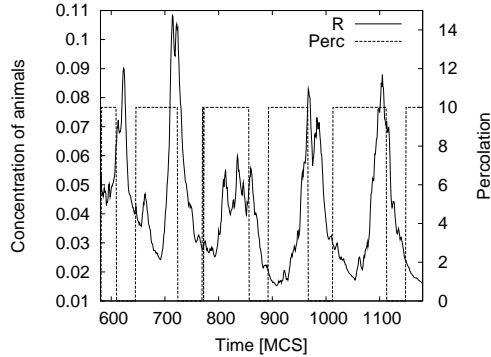} \
\caption{Time evolution of $R$  vs percolation intervals of $D$. To plot the intervals, in our simulations we assigned the value of 10 to each time step in which a detritus spanning cluster was detected, and 0 otherwise. Parameters of the simulation: $b_R=0.8$, $b_C=0.8$, $b_P=0.4$. \label{fig: r vs perc}}
\end{figure}

Our hitherto findings indicate that there could be a connection between the emergence of density waves and the percolation of detritus on the lattice. In order to verify this hypothesis we checked in every step of our simulations if there exists a cluster of detritus spanning the whole lattice. The results are shown in Fig.~\ref{fig: r vs perc}. We see that indeed there are percolation periods (indicated by the rectangles) separated by time intervals, in which no spanning clusters occur. Moreover, the peaks in the resource density accrue at the ends of the percolation intervals. The reason  is simply that after the emergence of a spanning cluster it takes some time for the wave to form and then to travel.
It should be mentioned that no spanning clusters were observed in case of a coexistence state or an almost monotonic decline (see Section~\ref{asymptotics} for details). In both cases in every time step the cluster distributions were similar to the left plot of Fig.~\ref{fig: d cluster dist}. It seems that the density blooms are possible only if the percolation of the detritus occurs in the system.

Percolating detritus clusters shown in Fig.~\ref{fig: r vs perc} may be interpreted as an accumulation of nutrients in the system. In ecology, a response of an ecosystem to increased levels of nutrients is known as eutrophication~\cite{mal2004}. High level of nutrients stimulates the primary production, causing a quick population increase (i.e. a bloom) of species such as algae in aquatic systems. These blooms may have many ecological effects, among others a decreased biodiversity~\cite{hor2002}, changes in species composition and dominance~\cite{ber2001} and toxicity effects~\cite{and1994}.

\subsection{Phase diagrams}
\label{phase diagrams}

To collect more information on the behavior of the system we performed a series of simulations with different sets of the model parameters. All birth rates of the species were varied from 0.1 to 0.9 with step 0.1. For each parameter set we run 100 independent runs. A set was tagged as an 'alive' one if at least one out of those 100 runs ended up in a coexisting state. Similarly, a set was labeled as a 'wave' one if at least in one of the runs the density waves were observed. 
The collection of all 'alive' sets constitutes the 'alive' phase. The remaining sets form the 'dead' phase, which is divided into two subphases: the (almost) 'monotonic' one and the 'wave' one (see Section~\ref{asymptotics} for more details). From our experience it follows that the transitions between phases are very sharp, i.e. if for a particular set of parameters one run ended in a coexistence state, then the most of the 100 runs led to the coexistence state as well.

\begin{figure}
\centering
\includegraphics[scale=0.6]{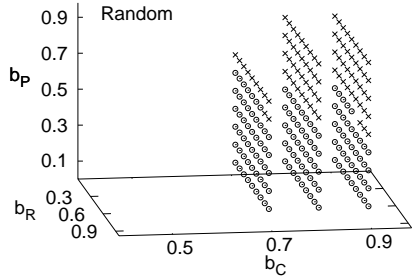}
\caption{A discrete 3D phase diagram in $(b_R,b_C,b_P)$ space obtained for $Random$ strategy. Circles correspond to the wave states, crosses are the alive states. All birth rates take values from 0.1 to 0.9, with the step 0.1. \label{fig: phase diag 3d (2)}}
\end{figure}

From the results of the above simulations one could generate a discrete 3D phase diagram in $(b_R,b_C,b_P)$ space. An example of such diagram is shown in Fig.~\ref{fig: phase diag 3d (2)} for the $Random$ strategy. However, since those diagrams are not really readable, we will analyse its 2D sections at fixed resource's birth rates instead.

\begin{figure}
\centering
\includegraphics[scale=0.5]{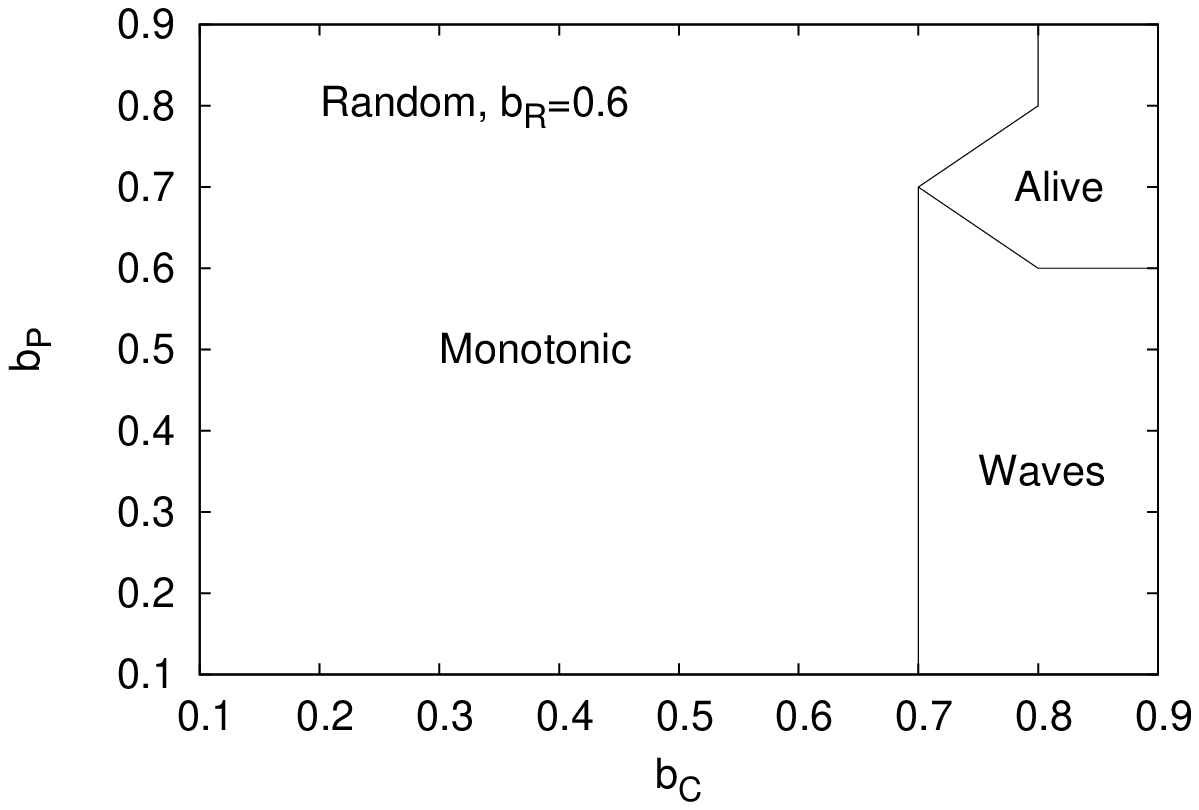}
\includegraphics[scale=0.5]{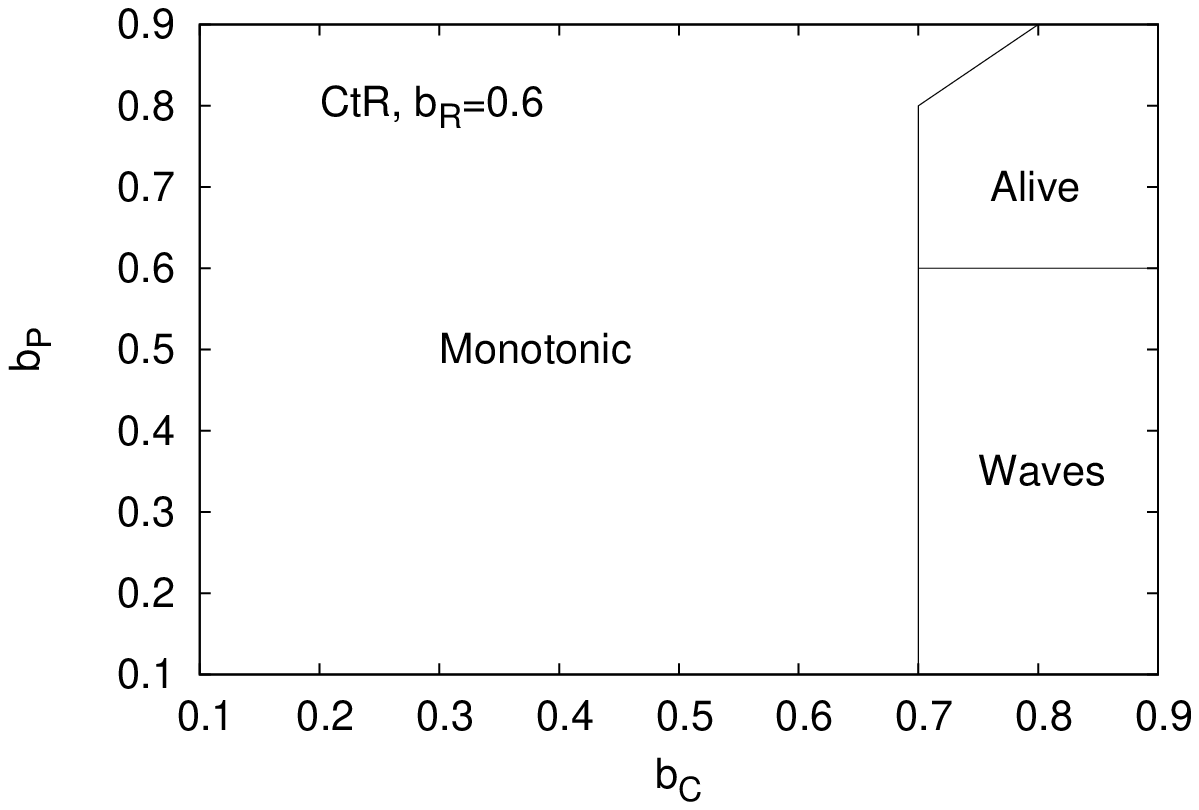}\\
\includegraphics[scale=0.5]{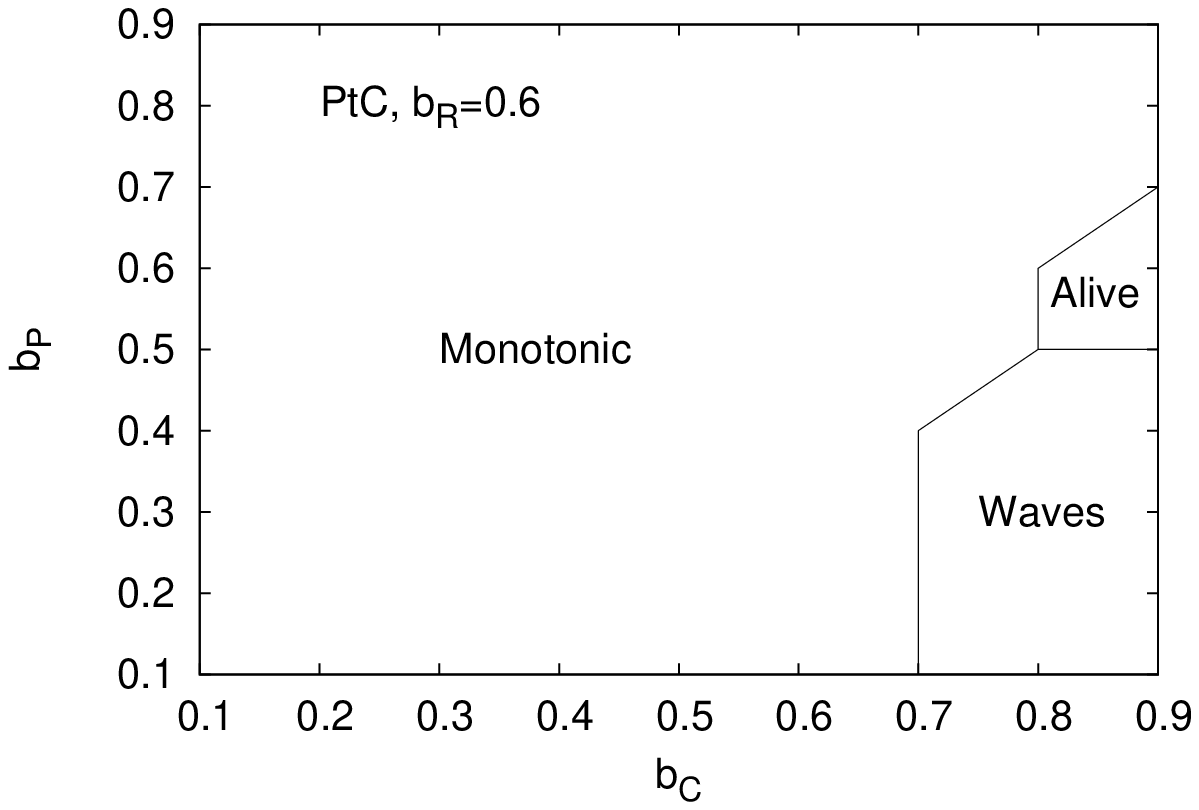}
\includegraphics[scale=0.5]{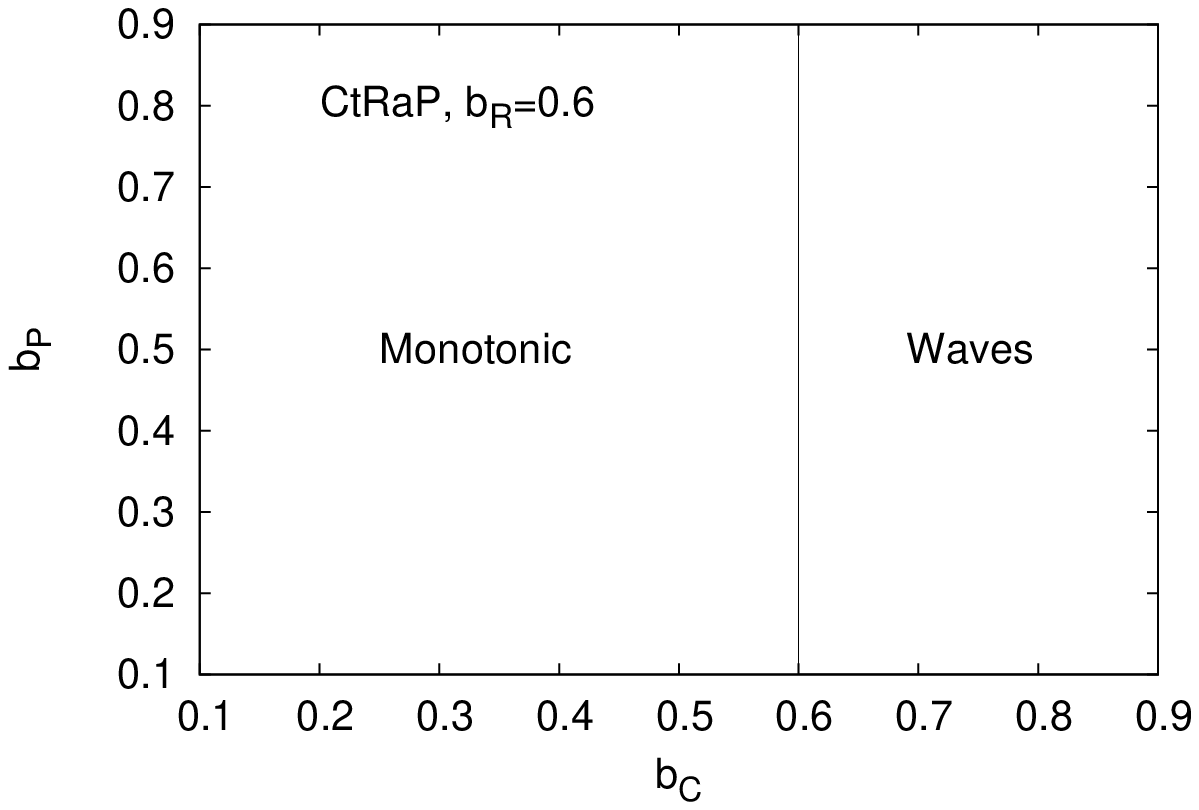}\\
\includegraphics[scale=0.5]{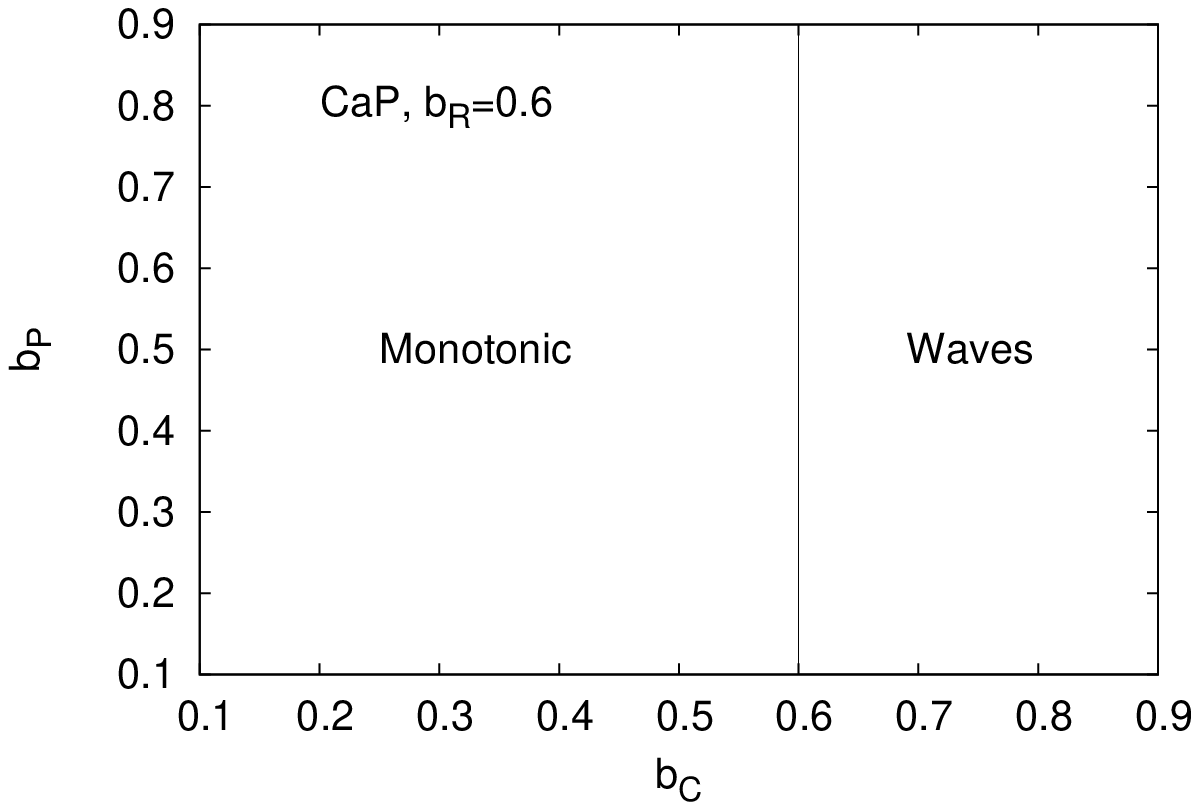}
\includegraphics[scale=0.5]{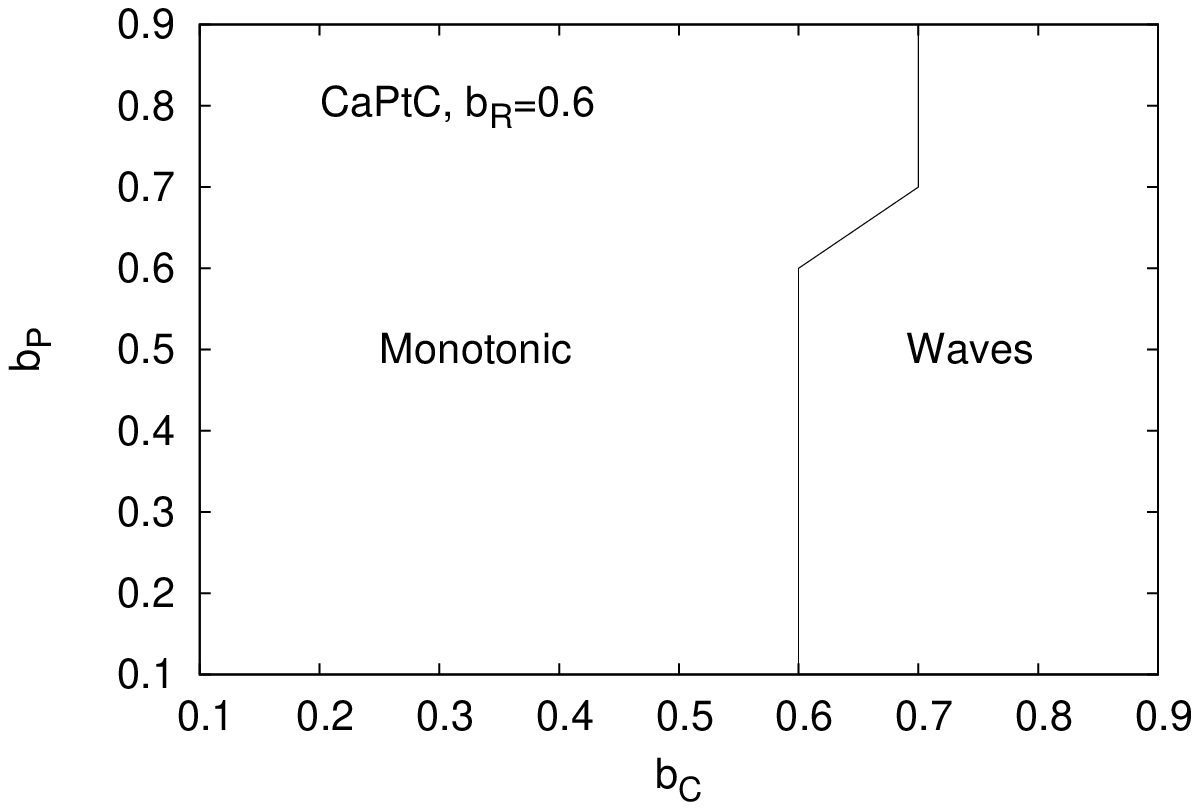}
\caption{Phase diagrams in $(b_C,b_P)$ plane at $b_R = 0.6$ for different proliferation strategies. Birth rates take values from 0.1 to 0.9, with the step 0.1. \label{fig: phase diag 2d}}
\end{figure}

In Fig.~\ref{fig: phase diag 2d}, 2D phase diagrams in $(b_C,b_P)$ planes at $b_r=0.6$ for different proliferation strategies are shown. For other values of $b_R$ we would get similar diagrams with only small changes in the areas of different phases. It is due to the fact that the viability of the system depends only weakly on $b_R$~\cite{jsw10,szw12a}.
We see that there is a critical value of $b_C$ for each proliferation strategy, above which the density waves appear in the system. In other words, only if the birth rate of the consumers is high enough, they are able to follow the resource's wave front according to the mechanism shown in Fig.~\ref{fig: cwaves}. Otherwise they will invade the lattice too slowly and the resources will die out before being caught up by the consumers. In this case we will observe after an initial density bloom the almost monotonic decline of the species densities towards the absorbing state (bottom left plot of Fig.~\ref{fig: asymptotics}).

Moreover, in the case of $Random$, $CtR$ and $PtR$ strategies, i.e. the strategies that support the 'alive' state as well, we also observe a critical value of $b_P$, below which the density blooms occur. Thus the emergence of the waves require the birth rate of predators to be small enough to allow an almost uncontrolled bloom of the consumers. If $b_P$ is higher than the critical value, the predators are able to suppress the quick increase of the consumers' density and the system arrives at a coexistence state. These results corroborate the existence of "top-down" interactions that impart stability to food webs~\cite{estes2011}.

\begin{figure}
\centering
\includegraphics[scale=0.5]{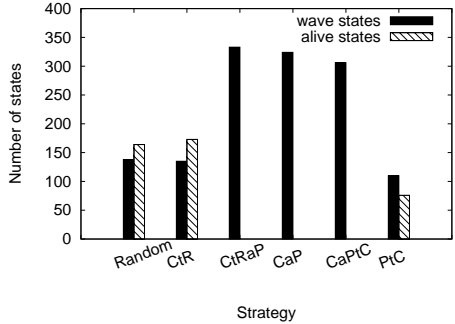} 
\caption{Both total number of density wave states and total number of alive states in the 3D phase diagrams as  functions of proliferation strategy.\label{fig: total wave and alive}}
\end{figure}

In order to provide more information on different proliferation strategies, we have counted both the 'alive' and 'wave' states in the discrete 3D phase diagrams as the one shown in Fig.~\ref{fig: phase diag 3d (2)}. The results are shown in Fig.~\ref{fig: total wave and alive}.
Again, we see that the proliferation strategies split into two groups - those which support only the two absorbing states ($CtRaP$, $CaP$ and $CaPtC$) and those which end up in a coexisting state as well ($Random$, $CtR$ and $PtC$). As far as the first group is concerned, the strategies are very similar to each other, since the volume of the 'wave' phase depends only little on the particular strategy. Among the strategies building up the other group, there is a strong similarity between $Random$ and $CtR$~\cite{szw12a}. While the volume of the 'alive' phase is slightly bigger for the $CtR$ strategy, we observe more 'wave' states in the phase diagram of the $Random$ one. The $PtC$ strategy differs from the other two in the size of the phases. However, the differences are smaller in case of the wave states.

\section{Conclusions}
\label{conclusions}

A simple spatial model of a three level food web with a closed nutrient cycle has been investigated. The results complement our previous findings on the stability of such a system~\cite{jsw10,szw12a}.

The time evolution of the model food web reveals two asymptotic states (Fig.~\ref{fig: asymptotics}): an absorbing one with all species being extinct, and a coexisting one, in which concentrations of all species are non-zero. Theoretically, one could expect that a state with species $R$ and $C$ having non-zero concentrations and $P$ being extinct is possible as well. Such a state was not observed in the simulations, because the predators are indispensable for the survival of the system. 

We found two possible ways for the system to reach the absorbing state depending on the particular values of the model parameters. In some cases the densities increase very quickly at the beginning of a simulation and then decline slowly and almost monotonically (small fluctuations disregarded). In others, well pronounced peaks in the $R$, $C$ and $D$ densities appear regularly before the extinction. 

Those peaks correspond to density waves traveling through the system, a phenomenon which is often observed in many complex systems in nature. Understanding the mechanism that triggers those waves was the focus of the present work. We have shown that several conditions have to be met for the waves to emerge:
\begin{enumerate}
\item An $RP$ node surrounded by the detritus is needed to initiate the resource wave.
\item A spanning cluster of the detritus on the lattice is required to feed the front of the resource wave.
\item The consumers must be able to reproduce quickly to follow the resource wave. In other words, their birth rate $b_C$ must be higher than a critical value.
\item The birth rate $b_P$ of the predators must be lower than a critical value to keep the number of the predators low and to allow almost uncontrolled grow of consumers followed by their quick decline due to natural causes.
\end{enumerate}
The birth rate $b_R$ of the resources and the different proliferation strategies turned out to be less important for the emergence of the density waves.

The existence of a critical value of $b_P$, above which the food web enters the coexistence phase (see Fig.~\ref{fig: phase diag 2d}) is consistent with the hypothesis of  top-down interactions being essential for the stability of food webs~\cite{estes2011}. Indeed, the emergence of the waves require the birth rate of predators to be small enough to allow an almost uncontrolled bloom of the consumers. If $b_P$ is higher than the critical value, the predators are able to suppress the quick increase of the consumers' density and the system survives.

The coupling between the percolation and spatio-temporal patterns has been already assumed or observed in many complex systems~\cite{mag2011,gre2001,kon2009,lip1999}. Since the percolation of the detritus in our model corresponds to an accumulation of nutrients, the processes we observe in the model reveal some resemblance to the eutrophication phenomena~\cite{mal2004} occuring in both aquatic and terrestrial ecosystems. Within our simple model this accumulation induces density blooms which drive the entire food web to extinction. In real ecosystems the outcome of eutrophication is usually not lethal for a system as a whole. Nevertheless, its ecological impact is severe, since it may result in loss of species, decreased biodiversity, changes in species composition and dominance or toxicity effects. 

Identifying percolation as one of the triggers for the density blooms may have a practical impact for the biological control. From the theory of percolation of disordered media~\cite{sta1994} it follows that it is enough to destroy some critical fraction of "transmission-promoting" sites to suppress the propagation of waves and their negative effects on the system. However, it should be checked which asymptotic state the system reaches after such an intervention.

\end{document}